# Effects of Paramagnetic Ferrocenium Cations on the Magnetic Properties of the Anionic Single-Molecule Magnet [Mn$_{12}$O$_{12}$(O$_2$CC$_6$F$_5$)$_{16}$(H$_2$O)$_4$]$^-$

Takayoshi Kuroda-Sowa,[1a†] Matthew Lam,[1b] Arnold L. Rheingold,[1b] Christoph Frommen,[1c] William M. Reiff,[1c] Motohiro Nakano,[1a‡] Jae Yoo,[1a] A. L. Maniero,[1d] Louis-Claude Brunel,[1d] George Christou,[1e,*] and David N. Hendrickson[1a,*]

*Department of Chemistry and Biochemistry-0358, University of California at San Diego, La Jolla, CA 92093-0358,*

*Department of Chemistry, University of Delaware, Newark, Delaware 19716, USA*

*Department of Chemistry, Northeastern University, Boston, MA 02115*

*Center for Interdisciplinary Magnetic Resonance, National High Magnetic Field Laboratory, Tallahassee, FL 32310*

*Department of Chemistry, Indiana University, Bloomington, IN 47405.*

## ABSTRACT

The preparation and physical characterization are reported for the single-molecule magnet salts [M(Cp')$_2$]$_n$[Mn$_{12}$O$_{12}$(O$_2$CC$_6$F$_5$)$_{16}$(H$_2$O)$_4$] (M = Fe, $n$ = 1, Cp' = C$_5$Me$_5$ (**2a**), C$_5$H$_5$ (**2b**), M = Co, $n$ = 1, Cp' = C$_5$Me$_5$ (**2c**), C$_5$H$_5$ (**2d**), M = Fe, $n$ = 2, Cp' = C$_5$Me$_5$ (**2e**), C$_5$H$_5$ (**2f**)) to investigate the effects of paramagnetic cations on the magnetization relaxation behavior of [Mn$_{12}$]$^-$ anionic single-molecule magnets. Complex **2a**•2H$_2$O crystallizes in the orthorhombic space group *Aba*2, with cell dimensions at 173 K of *a* = 25.6292(2), *b* = 25.4201(3), *c* = 29.1915(2) Å and *Z* = 4. Complex **2c**•2CH$_2$Cl$_2$•C$_6$H$_{14}$ crystallizes in the monoclinic space group *P*2$_1$/*c*, with cell dimensions at 173 K of *a* = 17.8332(6), *b* =

26.2661(9), $c$ = 36.0781(11) Å, $\beta$ = 92.8907(3)° and $Z$ = 4. These two salts consist of either paramagnetic $[Fe(C_5Me_5)_2]^+$ cations or diamagnetic $[Co(C_5Me_5)_2]^+$ cations, and $[Mn_{12}O_{12}(O_2CC_6F_5)_{16}(H_2O)_4]^-$ anions. The structures of the anions in the two salts are similar, consisting of a central $Mn_4O_4$ cubane moiety, surrounded by a nonplanar ring of eight Mn atoms that are bridged by and connected to the cube *via* $\mu_3$-$O^{2-}$ ions. The oxidation states of four Mn sites out of eight outer Mn ions in complex **2a** were assigned to be +2.75 from the valence bond sum analysis although the disordering of bridging carboxylates prevents more precise determination. On the other hand in complex **2c**, one Mn site out of eight outer Mn ions was identified as a $Mn^{II}$ ions, accommodating the "extra" electron; this was deduced by a valence bond sum analysis. Thus, the anion in complex **2a** has a $Mn^{II}_1Mn^{III}_7Mn^{IV}_4$ oxidation state description. The Jahn-Teller axes of the $Mn^{III}$ ions in both anions are roughly aligned in one direction. All complexes studied exhibit a single out-of-phase ac magnetic susceptibility ($\chi''_M$) signal in the 4.6 – 4.8 K range for complexes **2a-2d** and in the 2.8 - 2.9 K range for complexes **2e** and **2f** at 1 kHz ac frequency. The temperature of the $\chi''_M$ peaks is frequency dependent, as expected for single-molecule magnets. From Arrhenius plots of the frequency dependence of the temperature of the $\chi''_M$ maxima, the effective energy barriers $U_{eff}$ for changing spin from "up" to spin "down" were estimated to be 50 – 54 K for complexes **2a-2d** and 27 - 28 K for complexes **2e** and **2f**. The least-squares fits of the reduced magnetization data indicate both complexes **2a** and **2d** have ground states of $S$ = 21/2. High-frequency EPR spectra were recorded for complex **2a** at frequencies of 217, 327, and 434 GHz in the 4.5 - 30 K range. The observed transition fields were least-squares fit to give $g$ = 1.91, $D$ = -0.35 cm$^{-1}$, and $B_4^0$ = -3.6 $\times$ 10$^{-7}$ cm$^{-1}$ for the $S$ = 21/2 ground state. The effective energy barrier $U_{eff}$ is slightly lower than $U$ estimated from $D$, which is consistent with the thermally assisted tunneling model. Magnetization hysteresis loops were observed for complexes **2a** and **2c**. Although **2a** was oriented in a different manner as expected by strong magnetic field, both complexes show clear hysteresis loops with some steps on them, indicating the effect of the magnetic cation on the magnetization relaxation of the anionic $[Mn_{12}]^-$ complex is rather small. An 11 % $^{57}$Fe enriched complex **2b** was studied by means of Mössbauer spectroscopy down to as low as 1.7 K. Slow paramagnetic relaxation broadening and magnetic hyperfine splitting were evident in the low temperature spectra, indicating that the iron atoms feel a growing magnetic field owing to slow magnetization reversal in the $[Mn_{12}]^-$ anions.

**Introduction**

There is continuing interest in single-molecule magnets (SMM's).[2,3] The most extensively studied SMM is the dodecanuclear manganese complex $[Mn_{12}O_{12}(OAc)_{16}(H_2O)_4]$ (**1**) and its derivatives.[4-6] Due to their large-spin ground state ($S$=10) together with a uniaxial magnetic anisotropy (due to zero-field splitting $D\hat{S}_z^2$), each molecule has a potential barrier of $|D|S^2$ for reversal of the direction of its magnetization vector. This barrier leads to a magnetization hysteresis loop and the appearance of an out-of-phase signal in the ac susceptibility at low temperatures, where $k_B T$ is considerably smaller than the barrier height. The slow magnetization relaxation of a SMM may lead to them being used as the ultimate high-density memory device. From the entirely different point of view, SMM's are also attractive because they exhibit resonant quantum tunneling of magnetization (QTM) observed as steps in the magnetization hysteresis loops.[7-9]

Although it has been shown that these interesting properties arise from isolated molecules, it is also important to investigate how the environment around a [Mn$_{12}$] SMM affects them and *vice versa*. Recently, the magnetic properties of the salt [*m*-MPYNN][Mn$_{12}$O$_{12}$(O$_2$CPh)$_{16}$(H$_2$O)$_4$] have been reported by Awaga et al.[10] where [*m*-MPYNN]$^+$ is an organic cation radical, *m-N*-methylpyridinium nitronylnitroxide. On the basis of the X-band epr measurements and the observation of ac magnetization relaxation, they concluded that the organic radical with $S = 1/2$ enhances the magnetization relaxation of the [Mn$_{12}$]$^-$ complex and reduces the blocking temperature of the anion. However, it is well known that two relaxation processes are observed in the ac susceptibility response of several [Mn$_{12}$] complexes, the so called high-temperature (HT) and low-temperature (LT) relaxation processes.[10] The above-mentioned effects of the organic cation radical on the magnetization relaxation process of $[Mn_{12}]^-$ were interpreted[10] by suggesting that the magnetic field from the organic cation increased the magnetization relaxation for the $[Mn_{12}]^-$ SMM.

In order to investigate this point more thoroughly, salts of the singly reduced form of $[Mn_{12}O_{12}(O_2CC_6F_5)_{16}(H_2O)_4]$ (**2**) were prepared employing certain metallocenes. By comparing two high-temperature forms of $[Mn_{12}]^-$, we can determine the effect of a magnetic cation on the magnetization relaxation of $[Mn_{12}]^-$ complexes. We report the magnetic properties of the $[Mn_{12}]^-$ complexes with a paramagnetic $[Fe(C_5Me_5)_2]^+$ cations and with similar non-magnetic cations. $^{57}Fe$ Mössbauer spectroscopy is used to study the magnetization relaxation experienced by the ferrocenium cation as it is influenced by the oscillating magnetic field of the $[Mn_{12}]^-$ SMM.

## Experimental Section

**Compound Preparation.** All chemicals and solvents were used as received; all preparations and manipulations were performed under an argon atmosphere using Schlenk techniques. Bis(cyclopentadienyl)iron, bis(pentamethylcyclopentadienyl)iron, and bis(cyclopentadienyl)cobalt were purchased from Aldrich Chemical Co. Bis(pentamethylcyclopentadienyl)cobaltocenium hexafluorophosphate was purchased from Strem Chemicals and converted to $Co(C_5Me_5)_2$ by a literature procedure.[11] The $Mn_{12}$-acetate (**1**) was prepared by the literature method.[12] Neutral $[Mn_{12}O_{12}(O_2CC_6F_5)_{16}(H_2O)_4]$ (**2**) was prepared using the ligand substitution method described elsewhere.[13]

**$[Fe(C_5Me_5)_2][Mn_{12}O_{12}(O_2CC_6F_5)_{16}(H_2O)_4]\cdot2H_2O$ (2a$\cdot$2H$_2$O).** Complex **2** (100 mg, 0.023 mmol) was dissolved in $CH_2Cl_2$ (5 mL). An equivalent amount of bis(pentamethylcyclopentadienyl)iron (7.6 mg, 0.023 mmol) dissolved in $CH_2Cl_2$ (3 mL) was added to the above solution and stirred for 10 min. To the resultant solution, 10 mL of hexanes was added slowly. Several days later, brown-black plate crystals were obtained. The yield was 83 %. Anal. calcd for $C_{132}H_{42}F_{80}FeMn_{12}O_{50}$: C, 34.00; H, 0.91. Found: C, 33.71; H, 0.67.

**$[Fe(C_5H_5)_2][Mn_{12}O_{12}(O_2CC_6F_5)_{16}(H_2O)_4]$ (2b).** Brown powder obtained by the method described above showed a weak ac susceptibility signal, possibly indicating sample instability in solution. The following method was found to obtain a sample whose ac susceptibility signal is similar to the other one-electron reduced $[Mn_{12}]^-$ complexes. Complex **2** (100 mg, 0.023 mmol) was dissolved in $C_6H_6$ (5 mL). An equivalent amount of bis(cyclopentadienyl)iron (4.3 mg, 0.023 mmol) dissolved in $C_6H_6$ (3 mL) was added to the above solution and stirred for 10 min. The solution was evaporated by vacuum to dryness and the resultant brown powder was collected. The yield was ca. 90 %. The 11 % $^{57}$Fe-enriched sample for Mössbauer spectroscopy experiments was also obtained by the same method. Anal. calcd for $C_{140}H_{36}F_{80}FeMn_{12}O_{48}$ (as **2b**•$3C_6H_6$); C, 35.62; H, 0.77. Found: C, 36.20; H, 1.09.

**$[Co(C_5Me_5)_2][Mn_{12}O_{12}(O_2CC_6F_5)_{16}(H_2O)_4]$•$2CH_2Cl_2$•$C_6H_{14}$ (2c•$2CH_2Cl_2$•$C_6H_{14}$).** A method similar to that employed for the synthesis of **2a** was used. After standing for several days, brown-black needle crystals were obtained. The yield was 40 %. Anal. calcd for $C_{140}H_{56}Cl_4CoF_{80}Mn_{12}O_{48}$: C, 34.42; H, 1.16. Found: C, 34.01; H, 0.86.

**$[Co(C_5H_5)_2][Mn_{12}O_{12}(O_2CC_6F_5)_{16}(H_2O)_4]$ (2d).** The method used in the synthesis of **2a** was used. After standing for several days, brown-black plate crystals were obtained. The yield was 65 %. Anal. calcd for $C_{122}H_{18}CoF_{80}Mn_{12}O_{48}$: C, 32.64; H, 0.40. Found: C, 33.10; H, 0.35.

**$[Fe(C_5Me_5)_2]_2[Mn_{12}O_{12}(O_2CC_6F_5)_{16}(H_2O)_4]$ (2e).** Due to the same reason as for complex **2b**, a similar method was used for the preparation of complex **2e** except two equivalent amount of bis(pentamethylcyclopentadienyl)iron were used. Anal. calcd for $C_{172}H_{98}F_{80}Fe_3Mn_{12}O_{48}$ (as **2e**•$Fe(C_5Me_5)_2$); C, 39.13; H, 1.87. Found: C, 40.52; H, 1.92.

**[Fe(C_5H_5)_2]_2[Mn_{12}O_{12}(O_2CC_6F_5)_{16}(H_2O)_4] (2f).** The same method used for the preparation of complex **2b** was employed. Anal. calcd for $C_{152}H_{48}F_{80}Fe_4Mn_{12}O_{48}$ (as **2f**•$2Fe(C_5H_5)_2$); C, 36.19; H, 0.96. Found: C, 36.77; H, 1.08.

**X-ray crystal structure analyses.** Crystal data collection and refinement parameters for complexes **2a**•$2H_2O$ and **2c**•$2CH_2Cl_2$•$C_6H_{14}$ are given in Table 1. Suitable crystals for data collection were selected and mounted in nitrogen-flushed, thin-walled capillaries. Data were collected on a Siemens P4 diffractometer equipped with a SMART/CCD detector using Mo K$\alpha$ radiation ($\lambda$ = 0.710 73 Å). The systematic absences in the diffraction data were uniquely consistent with the reported space groups. The structures were solved using direct methods, completed by subsequent difference Fourier syntheses, and refined by full-matrix least-squares procedures. Empirical absorption corrections were applied to the data for complex **2c**•$2CH_2Cl_2$•$C_6H_{14}$ using SADABS. All non-hydrogen atoms were refined with anisotropic displacement coefficients. For complex **2c**•$2CH_2Cl_2$•$C_6H_{14}$, a part of hydrogen atoms of a hexane molecule were determined from difference Fourier analyses. All other hydrogen atoms in both structures were treated as idealized contributions. All software and sources of the scattering factors are contained in the SHELXTL (version 5.10) program library (G. Sheldrick, Siemens XRD, Madison, WI).

**Physical Measurements.** DC magnetic susceptibility data were collected on microcrystalline or a single-crystal sample restrained in eicosane to prevent torquing on a Quantum Design MPMS5 SQUID magnetometer equipped with a 5.5 T magnet. A diamagnetic correction to the observed susceptibilities was applied using Pascal's constants. Alternating current (ac) susceptibility measurements were carried out on a Quantum Design MPMS2 SQUID magnetometer equipped with a 1 T magnet and capable of achieving temperatures of 1.7 to 400 K. The magnitude of ac field is fixed to 0.1 mT, oscillating at a frequency in the range of 5 to 1000 Hz. Magnetization hysteresis loops were collected on a Quantum Design MPMS5 SQUID magnetometer employing oriented single crystals. The alignment of single crystals was carried out with eicosane while keeping the samples in a 5.0 T field at a temperature above the melting point (308-312 K) of eicosane for 15 min, after which the temperature was gradually decreased below the melting point to solidify the eicosane in order to constrain the sample. In this manner, one can measure hysteresis loops with the external magnetic field applied parallel to the easy axis of magnetization. For the case of a measurement with a magnetic field applied perpendicularly to the easy axis of magnetization, a sample prepared as described above was rotated 90° manually in the sample holder and fixed using a non-magnetic tape. Applied magnetic fields were varied from 2.5 T to −2.5 T to generate hysteresis loops.

High-frequency EPR measurements were performed at the National High-Magnetic Field Laboratory in Tallahassee, Florida. An Oxford superconducting magnet system (14.5 T) capable of high sweep rates (0.5 T/min) was used. The microwave source was a Gunn diode of 110 GHz nominal frequency. Harmonic generators permitted experiments at frequencies of $n \times 110$ GHz, where $n$ is an integer between 1 and roughly 4 as the power falls off rapidly at the higher harmonics. High-pass filters were used to cut out lower harmonics; however, higher harmonics frequencies were evident in the experimental spectra. A helium-cooled bolometer was used as the detector.

Zero-field $^{57}$Fe Mössbauer spectra were determined using a conventional constant acceleration spectrometer with a source of 50 MCi $^{57}$Co electroplated onto the surface and annealed into the body of a 6-μm-thick foil of high-purity rhodium in a hydrogen atmosphere. The details of cryogenics, temperature control, etc. have been described previously.[14]

**Results and Discussion**

**Structure of 2a•2H$_2$O.** This complex salt crystallizes in the orthorhombic *Aba*2 space group with four formula units in the unit cell. Figures 1 and 2 show the core structure of the [Mn$_{12}$]$^-$ anion and the molecular packing of **2a•2H$_2$O**, respectively. Decamethylferrocenium cations and [Mn$_{12}$]$^-$ anions are arranged in the quasi-NaCl-type structure as shown in Figure 2. Each component molecule has the two-fold symmetry axis parallel to the crystallographic *c*-axis. While the easy axis of magnetization of [Mn$_{12}$]$^-$ anion aligns with the crystallographic *c*-axis, the *g$_{//}$* axis of a decamethylferrocenium cation lies in the *a-b* plane. The structure of the [Mn$_{12}$]$^-$ anion in **2a•2H$_2$O** is quite similar in many respects to the previously characterized [Mn$_{12}$] complexes.[5,12,13,15-19] There is a central Mn$_4$O$_4$ cubane moiety, surrounded by a nonplanar ring of eight Mn atoms that are bridged by and connected to the cube *via* $\mu_3$-O$^{2-}$ ions. It should be noted that the structure of the [Mn$_{12}$]$^-$ anion in **2a** is described as a mixture of two different geometrical isomers involving different arrangements where the four H$_2$O ligands bond to different Mn$^{III}$ ions, the 1:1:1:1 form and the 2:2 form, as shown in Figure 1. The difference between them is only one carboxylic ligand out of eight. In the 1:1:1:1 form, the carboxylate ligand bridges between the Mn(5) and Mn(6) ions while in the 2:2 form it bridges between the Mn(5) and Mn(4) ions. Table 2 lists the average Mn-O bond lengths around each Mn ion. The average of the six Mn-O bond lengths for the Mn(1) and Mn(2) ions are quite short (1.91 Å). On the other hand, the six Mn-O bonds of the other Mn ions are divided into two categories: four shorter (1.93 – 1.99 Å, equatorial) and two longer (2.17 – 2.21 Å, axial) ones. These Mn-O bond lengths are strongly influenced by the oxidation states of each Mn ion, which will be discussed (*vide infra*) on the basis of the valence bond sum analysis. The elongation of axial Mn-O bond lengths compared to the equatorial ones is caused by the Jahn-Teller distortion in $d^4$ $Mn^{III}$ ions. Although the $R$-value is relatively high due to disordering of the carboxylate group, we can identify the Jahn-Teller axes. Four water oxygen atoms are involved in the elongated Mn-O bonds. The eight Jahn-Teller axes are roughly parallel to the crystallographic $c$-axis with angles of 15° - 40°. The alignment of the Jahn-Teller axes must play an important role in determining the magnitude of the molecular zero-field interaction $\mathbf{D}$ tensor[18,19] which results from coupling of the single-ion $\mathbf{D}$ tensors at each Mn ion. This is consistent with the observation of dominant high-temperature phase in the ac magnetization measurement.

In the previous paper on the one-electron reduced [$Mn_{12}$] complex of [$PPh_4$][$Mn_{12}O_{12}(O_2CEt)_{16}(H_2O)_4$],[13] the reduction site was assigned to one of $Mn^{III}$ ions on the basis of the bond valence sum analysis. A bond valence sum[20] is an empirical value, based on crystallographically determined metal-ligand bond distances, that may be used to determine the oxidation state of a metal. Bond valence sums($s$) were calculated using eq 1,

$$s = \exp[(r_0 - r)/B] \qquad (1)$$

where $r$ is the observed bond length, and $r_0$ and $B$ are empirically determined parameters. Values for $r_0$ are tabulated for $Mn^{n+}$ ($n = 2, 3, 4$).[20] Bond valence sum analysis has been used to verify oxidation states in metalloenzymes and superconductors,[21,22] and in [$Mn_{13}O_8(OEt)_6(O_2CPh)_{12}$],[23] a compound having Mn atoms in three different oxidation states [six $Mn^{II}$, six $Mn^{III}$, and one $Mn^{IV}$].

A bond valence sum analysis was carried out for **2a**•$2H_2O$ and the results are shown in Table 3. The assignment of +4 oxidation states for Mn(1) and Mn(2) and +3 oxidation states for Mn(3) and Mn(5) is reasonable. However, the calculated values of $s$ assuming +3 oxidation states for Mn(4) and Mn(6) are smaller than 3, indicating somewhat lower oxidation states than +3. Since these Mn ions are involved in the disordering of the bridging ligand, it is reasonable to assign +2.75 oxidation states for them. The additional electron is trapped on either these two Mn ions (four sites) depending on the disordering of the carboxylate.

**Structure of 2c•2CH$_2$Cl$_2$•C$_6$H$_{14}$.** This complex salt crystallizes in the monoclinic $P2_1/c$ space group. The structure shows a [Co(C$_5$Me$_5$)$_2$]$^+$ cation, a [Mn$_{12}$O$_{12}$(O$_2$CC$_6$F$_5$)$_{16}$(H$_2$O)$_4$]$^-$ anion, two CH$_2$Cl$_2$ molecules and one C$_6$H$_{14}$ molecule. Plots of the anion and a molecular arrangement in the unit cell are shown in Figures 3 and 4, respectively. As can be seen from Figure 3, the [Mn$_{12}$O$_{12}$(O$_2$CC)$_{16}$O$_4$] core structure is almost the same as that of **2a•**2H$_2$O, except for the water ligand arrangement. Four water molecules coordinate to the axial positions of three peripheral Mn ions with a 2:1:1 arrangement, where the Mn(11) atom is coordinated to two water molecules. The molecular packing in the unit cell is shown in Figure 4. The easy axes of magnetization of each [Mn$_{12}$]$^-$ anion are canted from the $b$-axis by 6.9°, and are also canted relative to each other by 13.8°.

The average Mn-O bond lengths are listed in Table 2 and the results of a bond valence sum analysis are shown in Table 3. It is reasonable to assign the four Mn ions in the central cubane framework [Mn(1) to Mn(4)] as having a +4 oxidation state and the remainder of the Mn ions, except for the Mn(11), as having a +3 oxidation state. The Mn(11) ion is special because the bond valence sum value is closer to 2, indicating a +2 oxidation state for this ion. We can conclude the extra electron is trapped at the Mn(11) ion to form the valence-trapped electronic structure of Mn$^{II}$Mn$^{III}$$_7$Mn$^{IV}$$_4$. This is the second example for [Mn$_{12}$]$^-$ anions to show clearly where the extra electron resides.[13] The reduction of one Mn$^{III}$ ion, rather than a core Mn$^{IV}$ ion, to a Mn$^{II}$ ion can be rationalized as following. For a Mn$^{IV}$ ion to become reduced, it must undergo significant structural rearrangement since a Jahn-Teller distortion is expected for the resulting high-spin Mn$^{III}$ ion. A Jahn-Teller distorted Mn$^{III}$ ion in the cubane core would create strain in the apparently rigid [Mn$_4$O$_4$] cubane unit. On the other hand, the reduction of an outer Mn atom of the ring, where the bonding framework is less rigid, concomitant with the additional equatorial elongation on one outer Mn atom would not perturb the basic [Mn$_{12}$O$_{12}$] structure significantly.

Each Jahn-Teller axis of seven Mn$^{III}$ ions is roughly parallel to each other to the same extent as found in **2a**•2H$_2$O. This is consistent with the fact that **2c**•2CH$_2$Cl$_2$•C$_6$H$_{14}$ also shows high-temperature phase dominantly in the ac magnetization measurement (*vide infra*).

**Reduced [Mn$_{12}$] Complexes Characterized by AC Magnetic Susceptibility.** Ferrocene is known to have an appropriate redox potential to reduce some [Mn$_{12}$] complexes.[13] Decamethylferrocene, cobaltocene, and decamethylcobaltocene as well as ferrocene were employed in one- and two-electron reductions of several [Mn$_{12}$] complexes. Reduction products were subjected to ac susceptibility measurements to investigate their magnetization relaxation characteristics. It is well known that a SMM exhibits a frequency-dependent out-of-phase ac susceptibility signal ($\chi''_M$). Figure 5 shows plots of $\chi'_M T$ vs $T$ (top) and $\chi''_M$ vs $T$ (bottom) for complex **2a**•2H$_2$O at various frequencies. Since one intense peak is observed in the 3 – 6 K range in the frequency range of 5 – 997 Hz, we can conclude that the HT relaxation phase has a dominant contribution to the magnetization relaxation process of **2a**•2H$_2$O. The weak $\chi''_M$ peak seen at ~2 K is probably due to a small amount of a second structural form of complex **2a** that exhibits the LT relaxation.

The observed frequency-dependent out-of-phase ac signals indicate that complex **2a**•2H$_2$O functions as a SMM. Figure 6 shows a double-well potential energy diagram for a $S$ = 21/2 SMM where one molecule reverses its magnetization direction from spin "up" to spin "down" in zero experimental magnetic field. The potential energy barrier $U$ is given by $U = |D|(S_z^2 - 1/4)$. At the $\chi''_M$ peak temperature, the rate of flip of the magnetic moment of a molecule between the "up" and "down" states is equal to the ac frequency.

The frequency dependence of the $\chi''_M$ peak temperature can be analyzed by means of the Arrhenius law. When the natural logarithm of the relaxation time $\tau$, evaluated from $1/(2\pi\nu)$ where $\nu$ is the ac frequency, is plotted versus the inverse of the $\chi''_M$ peak temperature $T$, the effective energy barrier $U_{\text{eff}}$ and pre-exponential factor $\tau_0$ can be evaluated with the following equation:[24]

$$\tau = \tau_0 \exp(\frac{U_{eff}}{k_B T}) \qquad (2)$$

, where $k_B$ is the Boltzmann constant. The Arrhenius plot for complex **2a•2H$_2$O** is shown in Figure 7, where the solid line shows the result of a least-squares fit of the ac susceptibility relaxation data to eq 2. For complex **2a•2H$_2$O**, the effective energy barrier $U_{\text{eff}}$ was determined to be 54 K with a pre-exponential factor of $2.1 \times 10^{-9}$ s. Frequency-dependent out-of phase ac signals were also seen for all the complexes studied, including the doubly reduced [Mn$_{12}$]$^{2-}$ dianions, which indicates they also function as SMM's. Table 4 lists the results of the ac susceptibility and relaxation fitting data for all of the complexes studied. [Mn$_{12}$] complexes studied here show $\chi''_M$ peaks at 5.9 K for neutral complex **2**, in the range of 4.6 - 4.8 K for the anionic complexes **2a − 2d**, and in the range of 2.8 - 2.9 K for the dianionic complexes **2e** and **2f**, respectively, at 1 kHz ac frequency. The ac out-of-phase signals of complexes **2**, **2a**, and **2e** are shown in Figure 8. The effective energy barriers $U_{\text{eff}}$ were estimated to be 50 − 54 K for the monoanions and 27 - 28 K for the dianions. The former values are comparable to those of anionic [Mn$_{12}$]$^-$ SMM's such as [PPh$_4$][Mn$_{12}$O$_{12}$(O$_2$CEt)$_{16}$(H$_2$O)$_4$] (60.2 K[25]) and [PPh$_4$][Mn$_{12}$O$_{12}$(O$_2$CPh)$_{16}$(H$_2$O)$_4$] (55 K,[10] 57.5 K[17]). The latter ones are somewhat smaller than 39 K estimated from $U = |D|S_z^2$ for a dianion [PPh$_4$]$_2$[Mn$_{12}$O$_{12}$(O$_2$CHCl$_2$)$_{16}$(H$_2$O)$_4$].[26]

Two points should be stressed here. One is the absence or very small contribution of the LT phase in the ac susceptibility measurements, which is usually observed together with the HT phase.[13,18,19] The high-temperature relaxation phase is seen dominantly, especially for the anionic [$Mn_{12}$] complexes, independent of the magnetism of the counter cations. This is in contrast to the result of [*m*-MPYNN][$Mn_{12}O_{12}(O_2CPh)_{16}(H_2O)_4$],[10] where the paramagnetic cation is purported to drastically influence the magnetization relaxation of the [$Mn_{12}$]⁻ anion. The sizes and shapes of counter ions and/or solvent molecules incorporated into the crystal structure probably strongly affect the alignment of Jahn-Teller axes of Mn(III) ions in a [$Mn_{12}$] complex, which is the most influential factor to determine whether it shows a HT or a LT phase. The second point is that certain combinations of [$Mn_{12}$] complexes and metallocenes give two-electron reduced [$Mn_{12}$]$^{2-}$ complexes.[26] The ac out-of-phase signals for the [$Mn_{12}$]$^{2-}$ species are clearly distinct from those of the one-electron reduced [$Mn_{12}$]⁻ species. Ac susceptibility measurements can aid in determining the reduction state of a given [$Mn_{12}$] SMM.

**DC Magnetization Studied as a Function of the Magnetic Field.** In order to determine the ground spin state and the magnitude of the zero-field splitting for [$Mn_{12}$] complexes, dc magnetization data were collected in the temperature range of 2.0 − 4.0 K and at external fields of 2.0, 3.0, 4.0, and 5.0 T for polycrystalline samples. To avoid torquing during measurements, samples were fixed into an eicosane wax in zero magnetic field before the measurements. The observed magnetization data plotted as reduced magnetization, $M/(N\mu_B)$, for complexes **2a**•$2H_2O$ and **2d** vs $H/T$ are given in Figure 9. A spin Hamiltonian including an isotropic Zeeman interaction and axial zero-field splitting ($DS_z^2$) was used to least-squares-fit the data by assuming that only the ground state is populated in the 2.0 − 4.0 K and 2.0 − 5.0 T ranges. The contribution of the paramagnetic [$Fe(C_5Me_5)_2$]⁺ cations ($S =$ 1/2, $g_{//}$ = 4.43, $g_{\perp}$ = 1.35) is also included for **2a•2H$_2$O**. The matrix was diagonalized on each cycle, and a powder average was calculated. The TIP was held fixed at $2400 \times 10^{-6}$ cm$^3$mol$^{-1}$.

The variable-field magnetization data for complex **2a•2H$_2$O** were fitted assuming a $S =$ 21/2 ground state with the parameters of $g$ = 1.92 and $D$ = -0.38 cm$^{-1}$. If we assume a $S =$ 19/2 ground state, the best fit parameters are $g$ = 2.12 and $D$ = -0.46 cm$^{-1}$. A $g$-value greater than $g_e$ (2.0023) is not acceptable for the [Mn$_{12}$] complexes consisting of mainly Mn$^{III}$ and Mn$^{IV}$ ions. Although the [Mn$_{12}$]$^-$ anions reported before[13,17,25,26] have been shown to have $S =$ 19/2 ground state, it is not so surprising for the [Mn$_{12}$]$^-$ anion in complex **2a•2H$_2$O** to have a $S =$ 21/2 ground state. Since a one-electron reduction changes one Mn$^{III}$ ion with $S$ = 2 to a Mn$^{II}$ ion with $S$ = 5/2, this leads to a simple vector-sum calculation that gives $(7 \times 2 + 5/2) - 4 \times 3/2 = 21/2$, if the Mn ion maintains similar magnetic interaction with the neighboring Mn ions. The establishment of the $S$ = 21/2 ground state for complex **2a•2H$_2$O** is discussed later in the HFEPR section (*vide infra*). A fit of the reduced magnetization data for complex **2d** also indicates a $S$ = 21/2 ground state, where the best fit parameters were found to be $g$ = 1.89 and $D$ = -0.36 cm$^{-1}$. Thus, the fitting of the variable-field magnetization data clearly indicates that both complexes **2a•2H$_2$O** and **2d** have $S$ = 21/2 ground states.

**High-Frequency EPR of Complex 2a•2H$_2$O.** With HFEPR data the total spin $S$ of the ground state and the values of the parameters $D$ and $g$ can be determined more precisely and uniquely than with bulk magnetization data. This technique is ideally suited for complexes that have appreciable zero-field splitting and/or large spin ground state.[27-31] The advantage of HFEPR over magnetization vs field data is that the ground state can be probed directly. Peaks observed in high-field HFEPR spectra are due to transitions between sublevels of the ground state. Magnetization vs field data reflect averages and are not nearly as discriminating in defining spin Hamiltonian parameters. In HFEPR spectra Boltzmann populations are reflected in the relative intensities of peaks. At low temperatures only a few sublevels of the ground state are populated, and therefore only a few peaks are seen in the HFEPR spectrum. As temperature is increased and more states become populated, more peaks are observed in the spectrum.

It is well known that due to the large magnetoanisotropy, microcrystals of a SMM readily orient in the field and the resulting HFEPR spectra can thus be viewed as pseudo-single-crystal spectra. In other words, if the z-axis of all the molecules in the crystal unit cell are parallel to one another, then all the molecules are aligned parallel to the direction of the external field $H$. This is the case for complex **2a•2H$_2$O**. The ground state of a SMM is split by axial zero-field splitting and the spin Hamiltonian can be expressed in its simplest form as given in eq 3:

$$\hat{H} = g\mu_B \hat{H} \cdot \hat{S} + D[\hat{S}_z^2 - \tfrac{1}{3} S(S+1)] \tag{3}$$

where $\mu_B$ is the Bohr magneton and $g$ is Lande's factor. The first term is the Zeeman term, and the second is the axial zero-field interaction term. Although the crystal lattice has an orthorhombic symmetry, the $E$ term expressing the transverse anisotropy can be neglected because a difference between the $a$- and $b$- lattice constants is less than 1 % and the [Mn$_{12}$]$^-$ anion in **2a•2H$_2$O** can be regarded as retaining a pseudo axial symmetry. The parameter $D$ gauges the axial zero-field splitting of the ground state. The energy of each $M_s$ sublevel of the ground state is given as $E = M_s g\mu_B H + D[M_s^2 - \tfrac{1}{3} S(S+1)]$. Therefore, the resonance field at which a transition occurs between the $M_s$ and $M_s + 1$ zero-field components of the ground state is given as,

$$H_r(M_s) = \frac{g_e}{g_\parallel}[H_0 - (2M_s + 1)D'] \tag{4}$$

where $H_0$ is a resonance field of a free electron at given frequency and $D' = \dfrac{D}{g_e \mu_B}$ for parallel transitions.

We have performed HFEPR measurements for microcrystalline sample of **2a**•$2H_2O$ at the frequencies of 217.23, 326.26, and 434.24 GHz in the 4.5 − 30 K range. Figure 10 shows the 326.26 GHz HFEPR spectra measured at 4.5 and 15 K. At 4.5 K only the lowest energy $M_s$ zero-field component of the ground state is thermally populated. At 15K several $M_s$ levels are populated and many other transitions are seen. Since the new transitions occur at higher fields than the most intense peak seen at 4.325 T in the 4.5 K spectrum, it can be concluded that the zero-field splitting parameter $D$ is negative. The lower intensity peaks appearing at fields less than 4.325 T must be part of other signals, probably due to molecules having slightly different $D$ values as a result of defects in the molecules or the presence of two (or more) isomeric forms of complex **2a**•$2H_2O$. These weak signals can be also seen in the 217.23 GHz and the 434.24 GHz spectra measured at 15 K as shown in Figure 11.

The resonance fields we have identified that belong to the main fine structure pattern for all three frequencies are summarized in Table 5. These were analyzed to determine the spin and zero-field splitting parameters of the ground state. In Figure 12 the resonance field values are plotted versus the $M_s$ value for each of the $M_s \rightarrow (M_s + 1)$ transitions. The data are divided into three groups because three different microwave frequencies were used. A least-squares fit of the data was carried out by using eq 4. The best fit was obtained by using the parameters $g_{//} = 1.906$, $D = -0.352$ cm$^{-1}$ and assuming a $S = 21/2$ ground state.

Since it has been shown that the quartic zero-field interaction term $B_4^0 \hat{O}_4^0$ plays an important role in some SMM's,[28,31-33] we also tried to fit the data with the following spin Hamiltonian containing $B_4^0 \hat{O}_4^0$.[34]

$$\hat{H} = g\mu_B \hat{H} \cdot \hat{S} + D[\hat{S}_z^2 - \tfrac{1}{3}S(S+1)] + B_4^0 \hat{O}_4^0 \qquad (5)$$

where $\hat{O}_4^0 = 35\hat{S}_z^4 - 30S(S+1)\hat{S}_z^2 + 25\hat{S}_z^2 - 6S(S+1)$. Since we are in the high-field limit, transitions will occur at the following field values:

$$H_r(M_s) = \frac{g_e}{g_\parallel}[H_0 - (2M_s+1)(D' + 25B_4^{0'} - 30S(S+1)B_4^{0'}) - 35B_4^{0'}(4M_s^3 + 6M_s^2 + 4M_s + 1)] \qquad (6)$$

where $B_4^{0'} = \dfrac{B_4^0}{g_e \mu_B}$. The lines in Figure 12 result from a least-squares fitting of the resonance fields to eq 6. The best fit occurred for the following parameters: $S = 21/2$, $g_{\parallel} = 1.908$, $D = -0.351$ cm$^{-1}$ and $B_4^0 = -3.6 \times 10^{-7}$ cm$^{-1}$, which are essentially the same as those obtained by using eq 4. A negligibly small contribution of $B_4^0$ is reasonable for the almost straight lines in Figure 12. These parameters are comparable to those reported for other [Mn$_{12}$] SMM's.[17,34] If we assume the ground state spin of $S = 19/2$, the best-fit parameters are $g_{\parallel} = 2.03$, $D = -0.38$ cm$^{-1}$ and $B_4^0 = 1.7 \times 10^{-6}$ cm$^{-1}$, which can be excluded because the $g$-value is greater than $g_e$. The characterization of the ground state of complex **2a**•2H$_2$O obtained by fitting the variable-field magnetization data ($S = 21/2$, $g = 1.92$, $D = -0.38$ cm$^{-1}$) is in good agreement with the HFEPR results.

From the $D$-value of $-0.35$ cm$^{-1}$ ($= -0.51$ K) obtained by HFEPR data for complex **2a**•2H$_2$O, we can estimate the barrier height of the double well potential $U$ in Figure 6 to be 55.5 K, which is to be compared to the effective energy barrier $U_{\text{eff}}$ of 54 K obtained from the Arrhenius plot of the frequency dependence of the $\chi''_M$ peak temperature. The slight reduction of $U_{\text{eff}}$ compared to $U$ is consistent with the thermally assisted tunneling model.[35] In this model, the system initially in the left well is thermally activated to a fast-tunneling level near the top of the barrier *via* an Orbach phonon process, tunnels through the barrier and then spontaneously decays to the ground state in the right well. On resonance, this process then produces an effective reduction in the energy barrier. From the energy difference between $U_{eff}$ and $U$, we can say that tunneling might occur between $M_s = \pm 3/2$ or $\pm 5/2$ levels. The estimate of $U$ gives another confirmation of $S = 21/2$. If we assume $S = 19/2$, $U$ becomes to 49 K, which is smaller than $U_{eff}$ and inconsistent with our model.

**Magnetization Hysteresis Loops.** Recently, magnetization hysteresis loops ($M$ vs $H$) have been reported for SMM's including several [Mn$_{12}$] complexes,[7,15,16,25,30,32,36,37] [Fe$_8$O$_2$(OH)$_{12}$(tacn)$_6$]$^{8+}$,[38] (tacn = triazacyclononane) and [Mn$_4$O$_3$Cl(O$_2$CCH$_3$)$_3$(dbm)$_3$][39] (dbm$^-$ is the dibenzoylmethane monoanion). Steps are seen in the hysteresis loop at constant intervals of field. These steps in magnetization are due to a sudden increase in the decay rate of magnetization occurring at a specific magnetic field values and have been attributed to field-tuned quantum tunneling of the direction of magnetization. The observation of quantum tunneling of magnetization in complexes with not only integer spin but also half-integer spin indicates a transverse internal field plays an important role in the tunneling process.[32,39]

The magnetization hysteresis loops were determined for [Mn$_{12}$]$^-$ complexes with either paramagnetic or diamagnetic counter cations. Figure 13 shows the magnetization hysteresis loop observed at 1.85 K for microcrystals (1.5 mg) of complex **2c**•2CH$_2$Cl$_2$•C$_6$H$_{14}$, a salt with a diamagnetic cation. The hysteresis loop has similar features to that reported for [PPh$_4$][Mn$_{12}$O$_{12}$(O$_2$CEt)$_{16}$(H$_2$O)$_4$]:[25] the magnetization saturates at over 1 T to around 16 $N_B$ and shows hysteresis with a coercive field of 0.41 T. In the lower part of Figure 13 is shown the first derivative of the hysteresis plot. As the field is decreased from +2 T, the first step can be seen at zero field, followed by steps at –0.42, and –0.81 T.

On the other hand, the magnetization behavior of complex **2a**•2H$_2$O, a salt that has the paramagnetic decamethylferrocenium cation, is not straightforward. A 1.7 mg rectangular plate crystal of complex **2a**•2H$_2$O was oriented in a 5 T magnetic field at 350 K in a capsule with eicosane and cooled to fix it in space. The crystal was oriented so that the longest side of the crystal is parallel to the field. The field dependence of the dc magnetization measured with this orientation of the single crystal at 1.85 K is shown in Figure 14 with open circles. Although hysteresis can be seen with a coercive field of 0.34 T, the magnetization is not saturated up to 2.5 T. To our knowledge this is quite unusual for [$Mn_{12}$] SMM's. The non-saturated hysteresis loop implies the magnetization easy axis of [$Mn_{12}$]$^-$ anion was not parallel to the external field, being different from those before.[15,16,25] The sample was then rotated 90° so that the field applied normally to the flat plane of the crystal. The observed hysteresis loop at the same temperature is shown in Figure 14 with filled squares. The hysteresis loop shows clear saturation at over 1 T to *ca.* 20 $N\mu_B$ and a coercive field of 0.34 T. In the lower part of Figure 14, which shows the first derivative of the hysteresis plot obtained after rotation of the crystal by 90°, steps in the hysteresis loop can be observed at -0.04, -0.35, and -0.73 T. The steps correspond to increases in the rate of change of the magnetization, and are attributable to resonant tunneling between the quantum spin states. With the spin Hamiltonian of eq 3, the step interval value, $\Delta H$, in a hysteresis loop is given as $\Delta H = -nD/(g\mu_B)$, where $n$ = 0, 1, 2, … Using $D$ and $g$ obtained by HFEPR measurement, we can estimate a step interval $\Delta H$ to be 0.39 T, which is comparable to the observed value of 0.35 T. Although there are small differences in step interval $\Delta H$ or coercive fields between the complexes **2a**•$H_2O$ and **2c**•$2CH_2Cl_2$•$C_6H_{14}$, the effect of magnetic moment originating from the $Fe^{III}$ ion ($S$ = 1/2) on the magnetization relaxation of the [$Mn_{12}$]$^-$ complex is rather small as indicated by these magnetization hysteresis measurements.

As for the sample orientation of **2a**•$H_2O$ in the magnetic field, the above observations are fully consistent with the crystal structure analysis as shown in Figure 2. The easy axis of magnetization of the crystal at 350 K is perpendicular to that of the [$Mn_{12}$]$^-$ anion. Since the principal axis of the $g$-tensor of [$Fe(C_5Me_5)_2$]$^+$ cation is perpendicular to the easy axis of magnetization of the [$Mn_{12}$]$^-$ anion, the crystal was aligned by the magnetization anisotropy of the $[Fe(C_5Me_5)_2]^+$ cation, which comes from the anisotropy of the $g$ –tensor ($g_{//} = 4.43$ and $g_\perp = 1.35$). On the other hand, the magnetization anisotropy of $[Mn_{12}]^-$ anion comes from a zero-field splitting interaction, $DS_z^2$. Although the anisotropy energy of the latter with $S = 21/2$ is larger than the former, the effective $S$ value for the $[Mn_{12}]^-$ anion will be decrease with increasing temperature as exemplified by the $\chi_M T$-$T$ plot in the literature,[13] due to the thermal population of the excited states with lower $S$ values. Thus the anisotropy energy of the crystal is dominated by the anisotropy of $[Fe(C_5Me_5)_2]^+$ at 350 K.

**Mössbauer spectra of 2b.** The highly characteristic, relatively narrow singlet nuclear gamma resonance of the ferrocenium cation is an ideal probe of the low temperature relaxation dynamics of neighboring $[Mn_{12}]^-$ anions in a salt with the composition $[Fe(C_5H_5)_2]^+\bullet[Mn_{12}]^-$. The sample of complex **2b** was prepared enriched by 11 % $^{57}$Fe. The resulting $^{57}$Fe Mössbauer spectra are shown in Figure 15. First of all, the observation of a singlet Mössbauer spectrum at ambient temperature and 77 K (Figure 15) gives clear, additional confirmation of the one electron reduction of the $[Mn_{12}]$ cluster by ferrocene. That is, ferrocene exhibits a well-resolved quadrupole doublet spectrum with a splitting of ~2.4 mm/s.[40] There is no evidence of this low spin ferrous doublet (indicating residual unoxidized ferrocene) in the spectra. Such an impurity would be obvious in the present $^{57}$Fe enriched samples. The requirement for this level of enrichment was dictated largely by the high molar mass (4486 g/mol) of complex **2b**. In the context of the goal of obtaining acceptable signal to noise Mössbauer spectra for reasonable experimental counting times, study of the dilute natural abundance complex salt would be problematic save for the most intense of $^{57}$Co gamma ray sources. Typical unenriched ferrocenium salts with diamagnetic counter anions, e.g. $BF_4^-$, $PF_6^-$; $I_3^-$ etc. exhibit line widths in the range 0.3 to 0.4 mm/s for thin absorbers. However, the present material has a line width of ~0.6 mm/s at ambient temperature, which increases to ~1 mm/s at 77 K. We attribute the former (in part) to the enriched, relatively thick samples used in our experiments. The latter temperature dependence is likely the result of slow paramagnetic relaxation in the first few excited (zero field split) total spin states of the $[Mn_{12}]^-$ clusters.

At even lower temperatures, the ground state of the manganese cluster is essentially exclusively populated. One expects the longest relaxation times for the ground Kramers doublet of this spin manifold owing to the large angular momentum changes involved. Coincidentally, one observes partially resolved magnetic hyperfine splitting of the Mössbauer spectra with the onset of slow relaxation at the $Fe^{III}$ sites. However, the limiting spectral line shape is not that expected for the infinitely slow relaxation limit.[41] These observations for the low spin $Fe^{III}$ are explicable in terms of the negative zero field splitting of the ground spin state of $[Mn_{12}]^-$ ultimately leading to slow *intracluster* relaxation and *an internal field*. The following more general considerations are also important. Unlike high spin $Fe^{III}$, spin-lattice relaxation is an important and often dominant additional relaxation mechanism in the case of low spin $Fe^{III}$. Thus, one needs (I) *low temperatures* to vitiate the effects of phonons and lengthen spin-lattice relaxation times. One also requires (II) a degree of magnetic dilution ($Fe^{III}$-$Fe^{III}$ distances greater than ~ 8 Å) thus lengthening spin-spin relaxation times.[42] Finally, a source of small to moderate local magnetic fields that induce small electron Zeeman splittings further enhances the slow paramagnetic relaxation effect *via* magnetic polarization with decreasing temperature. (Recall that the (nominal) $S = 1/2$ ground state of the ferrocenium cation corresponds to a ***rapidly relaxing*** Kramers doublet whose degeneracy can only be removed by the presence of a magnetic field.) Of course the internal fields arising from the latter slow (electron) paramagnetic relaxation of $Fe^{III}$ lead to the observed nuclear Zeeman spitting of the low temperature Mössbauer spectra. Needless to say, the specific experimental conditions under which the Mössbauer spectra were determined largely satisfy (I). In the context of (II), this system is rather self-dilute vis a vis $Fe^{III}$-$Fe^{III}$ distances since the shortest such distance is ~ 18 Å. The pertinent cluster parameters, $U_{eff}$ ~ 50 K and $D$ ~ -0.5 K for the present mono-anionic manganese clusters are similar to those found previously[43,44] for the neutral, $S$ = 10 analogue. Thus, with decreasing $T$ and the onset of slow cluster magnetization reversal, one expects the development of substantial *internal* fields in the "single-molecule magnet ($[Mn_{12}]^-$) units". Somewhere above the blocking temperature for the latter magnetization reversal (~5 K), the cations will in turn commence to experience the growth of fluctuating *dipolar* fields owing to their $[Mn_{12}]^-$ neighbors. Finally (below ~ 2 K for $[Mn_{12}]$) the reversal enters a non-thermally assisted, temperature independent regime. The magnetization reversal now occurs *via* direct quantum tunneling[45] through the ~ 50 K energy barrier between the lowest, degenerate states of the ground Kramers doublet of the anion. In this situation, the ferrocenium cations are in the magnetic field environment presented by the random array of essentially equal numbers[46] of neighboring spin-up and spin-down $[Mn_{12}]^-$ clusters. *The limiting magnetic field experienced by a specific ferric cation in the vicinity of a given set of neighboring anions can be viewed as basically static on an iron-57 Mössbauer spectroscopy time scale*. This is consistent with the fact that the latter corresponds to ~100 ns while the cluster magnetization reversal time is of the order 2 months[47] for $[Mn_{12}]$ below ~ 2 K and is expected to be similar for the present mono anion.

Theoretical studies[48,49] suggest that a *distribution* of dipolar fields arises from the non-reversal of magnetization of the neutral $[Mn_{12}]$ cluster at low temperatures. These fields range to as large as ~ 600 Gauss with the maximum in the distribution centered near 250 Gauss. The exact field experienced at any given ferrocenium site from this source is a sensitive function or geometric considerations, particularly distance since such fields are proportional to $R^{-3}$. In any event, magnetic fields of this general magnitude are sufficient to decouple[50] the hyperfine interactions of the nuclear magnetic moments of the $I = 1/2$ and $I = 3/2$ states of $^{57}$Fe with the $S = 1/2$ of the cation and thus simplify the Mössbauer spectra and (ii) induce varying degrees of slow paramagnetic relaxation.[51] However at 3.3 K, residual rapidly relaxing [Fe(C$_5$H$_5$)$_2$]$^+$ is still evident just above zero velocity (Figure 15) while at 1.7 K this phase has largely disappeared. At lower temperatures, one may see achievement of the infinitely slow relaxation limit line shape as one approaches paramagnetic saturation with increasing $H/T$ at the ferrocenium sites. In this context, experiments in the helium-3 temperature range (1.6 to ~0.3 K) will be helpful and are planned. In passing, it then seems reasonable to attribute the broadness of the observed resonances to the aforementioned distribution of dipolar fields and thus a concomitant distribution of $^{57}$Fe relaxation times. Judging by the spectra and comparisons to the available literature;[41,52] these range from ~$10^{-9}$ to ~$10^{-8}$ sec.

The magnitude of the internal field at the [Fe(C$_5$H$_5$)$_2$]$^+$ sites is quite large, averaging some 35 T at 1.7 K, and is worthy of comment. Although detailed X-band esr measurements have not been made for the present $^{57}$Fe enriched system, internal hyperfine fields ($H_N$) of ~30 to as large as 44 T are explained by consideration (a) of the explicit expression for $H_N$ and (b) the range of $g$-factor anisotropy typical of ferrocenium systems. Specifically: $H_N = H_F + H_D + H_L$ where $H_F$ is the Fermi contact contribution which for low spin ferric, including covalency delocalization-reduction effects, is estimated at ~ - 10 T.[40,53] The dipolar term ($H_D$) scales as the electric field gradient tensor and is thus expected to be ~ 0 for the case of near zero quadrupole splitting effects generally characteristic of ferrocenium systems. Finally, there is the orbital contribution ($H_L$) where: $H_L = -2\mu_B < r^{-3} >_{3d} (g - 2) <S_Z>$. The esr spectra of ferrocenium systems usually exhibit axial symmetry with $g_{/\!/}$ ranging[54] between ~3.80 to ~ 4.40 for a mid-range of 4.10. Values of $g_\perp$ typically fall between ~ 1.20 to ~ 1.90 for an average of ~ 1.55. Hence, using appropriate values in the foregoing expression for $H_L$ and $S_Z$ = 1/2, one calculates a range of $H_L$ varying from ~ -12 T to ~ +47 T. When considered in conjunction with $H_F$ this corresponds to $H_N$ spanning −22 T to +37 T. It is clear from the expression for $H_N$ that the latter (positive) value of $H_L$ (correlating with $g_{\parallel}$) will lead to the large absolute values (30 to 40 T) observed for the hyperfine fields of the Mössbauer spectra of many magnetically ordered or infinitely slowly relaxing (paramagnetic) ferrocenium compounds.[55] In this context, applied field Mössbauer spectra of ferrocenium systems[56] and relevant ferricyanides[57] show that $H_N$ is indeed positive for such complexes.

The present Mössbauer spectra contrast strongly with the low temperature results for "Fe$_8$". This material is also SMM that exhibits slow magnetization reversal and ultimate quantum tunneling[38] with: $S = 10$, $U_{eff}$ ~25 K, $D$ = -0.27 K, a crossover temperature, $T_C$, to the tunneling regime of 0.7 K, and magnetization relaxation time in this regime of some $10^4$ sec (~3 hours). However, its Mössbauer spectrum[58] at 2 K corresponds to three, relatively well resolved, *narrow line width*, hyperfine patterns, reminiscent of bulk magnetic ordering although the material is not in fact ordered. In this case, each of the three crystallographicaly different sites[59] of the "Fe$_8$ " unit experiences *a single effective intra-cluster internal hyperfine field* as opposed to the distribution of (dipolar) fields about slowly relaxing $[Fe(C_5H_5)_2]^+$ cations considered herein.

**Concluding Remarks**

The results presented show that the effects of the paramagnetic metallocene cations on the magnetic properties of the $[Mn_{12}]^-$ SMM are negligibly small. This is in contrast to a result reported by Awaga et al.[10] They concluded that the organic radical cation enhances the magnetization relaxation of the $[Mn_{12}]$-benzoate anion. However, on the basis of the results presented here, it is not the magnetic cations that influence the magnetic relaxation properties of $[Mn_{12}]^-$ anion. This is probably due to the influence of the alignment of the Jahn-Teller axes of peripheral $Mn^{III}$ ions on the blocking temperature. In a recent paper,[19] we have shown in detail that Jahn-Teller isomers exist for S = 10 $Mn_{12}$ complexes. In one isomer the Jahn-Teller elongation axes of the 8 $Mn^{III}$ ions are nearly parallel. In the other isomer one or more of these Jahn-Teller elongation axes are oriented approximately perpendicular to the others in the $Mn_{12}$ molecule. This second isomer has a lower molecular symmetry and the resultant increase in transverse interaction terms [such as E $(\hat{S}_x^2 - \hat{S}_y^2)$] in the spin Hamiltonian lead to a much faster rate of tunneling of the magnetization for this second type of Jahn-Teller isomer. Thus, the second type shows $\chi''_M$ peaks in the 2-3 K range, whereas the first type of Jahn-Teller isomer exhibits $\chi''_M$ peaks in the 4-7 K range. This same type of Jahn-Teller isomerism is likely present in the $[Mn_{12}]^-$ SMM"s.

**Acknowledgment.** This work was supported by NSF grants (D.N.H. and G.C.). The ac susceptibility measurements were performed with an AC SQUID susceptometer provided by the Center for Interface and Material Science, funded by the W. A. Keck Foundation.

**Supporting Information Available:** X-ray structural data for complexes **2a**•$2H_2O$ and **2c**•$2CH_2Cl_2$•$C_6H_{14}$, consisting of crystallographic parameters, atomic coordinates, bond lengths and angles, and anisotropic thermal parameter. An X-ray crystallographic file, in CIF format, is also available.

**References.**

**Table 1.** X-ray Diffraction Data for **2a**•2H$_2$O and **2c**•2CH$_2$Cl$_2$•C$_6$H$_{14}$

|  | **2a**•2H$_2$O | **2c**•2CH$_2$Cl$_2$•C$_6$H$_{14}$ |
|---|---|---|
| formula | C$_{132}$H$_{42}$F$_{80}$FeMn$_{12}$O$_{50}$ | C$_{140}$H$_{56}$Cl$_4$CoF$_{80}$Mn$_{12}$O$_{48}$ |
| formula weight | 4662.79 | 4885.86 |
| crystal dimensions (mm) | 0.40 x 0.30 x 0.21 | 0.30 x 0.15 x 0.02 |
| crystal color | dark brown | brown |
| temperature (K) | 173(2) | 173(2) |
| crystal system | Orthorhombic | Monoclinic |
| space group | *Aba*2 | *P*2$_1$/*c* |
| *a* (Å) | 25.6292(2) | 17.8332(6) |
| *b* (Å) | 25.4201(3) | 26.2661(9) |
| *c* (Å) | 29.1915(2) | 36.0781(11) |
| *β* (deg) |  | 92.8907(3) |
| *V* (Å$^3$) | 19018.2(3) | 16877.7(15) |
| *Z* | 4 | 4 |
| *D*$_{calcd}$ (g/cm$^3$) | 1.628 | 1.923 |
| diffractometer | CCD area detector | CCD area detector |
| *μ* (mm$^{-1}$) | 9.89 | 11.92 |
| refinement, *wR*2 [a] | 0.3373 | 0.3264 |
| refinement, *R*1 | 0.1342 | 0.1569 |
| goodness of fit on *F*$^2$ | 2.753 | 1.828 |

[a] $wR2 = [\Sigma[w(F_o^2 - F_c^2)^2]/[\Sigma[w(F_o^2)^2]]^{1/2}$.

**Table 2.** Average Mn-O Bond Lengths (Å) in **2a**•2H$_2$O and **2c**•2CH$_2$Cl$_2$•C$_6$H$_{14}$

| Mn atom | methods of average | | |
|---|---|---|---|
| | 6 Mn-O[a] | 4 Mn-O[b] | 2 Mn-O[c] |
| | **2a**•2H$_2$O | | |
| Mn(1) | 1.91(4) | | |
| Mn(2) | 1.91(4) | | |
| Mn(3) | | 1.94(4) | 2.18(3) |
| Mn(4) | | 1.98(3) | 2.17(1) |
| Mn(5) | | 1.93(3) | 2.21(1) |
| Mn(6) | | 1.99(2) | 2.20(1) |
| | **2c**•2CH$_2$Cl$_2$•C$_6$H$_{14}$ | | |
| Mn(1) | 1.90(1) | | |
| Mn(2) | 1.89(2) | | |
| Mn(3) | 1.91(4) | | |
| Mn(4) | 1.90(4) | | |
| Mn(5) | | 1.92(4) | 2.22(1) |
| Mn(6) | | 1.93(5) | 2.20(1) |
| Mn(7) | | 1.92(5) | 2.20(2) |
| Mn(8) | | 1.93(3) | 2.21(5) |
| Mn(9) | | 1.93(7) | 2.16(5) |
| Mn(10) | | 1.94(6) | 2.15(1) |
| Mn(11) | | 2.12(2) | 2.21(1) |
| Mn(12) | | 1.91(5) | 2.13(6) |

[a]Average of six Mn-O bond lengths. [b]Average of four shorter (equatorial) Mn-O bond lengths. [c]Average of two longer (axial) Mn-O bond lengths.

**Table 3.** Bond Valence Sums for Each Mn Atom in Complexes **2a**•$2H_2O$ and **2c**•$2CH_2Cl_2$•$C_6H_{14}$ Assuming Various Charges[a]

| atom | **2a**•$2H_2O$ | | | **2c**•$2CH_2Cl_2$•$C_6H_{14}$ | | |
|------|-------|-------|-------|-------|-------|-------|
| | $Mn^{2+}$ | $Mn^{3+}$ | $Mn^{4+}$ | $Mn^{2+}$ | $Mn^{3+}$ | $Mn^{4+}$ |
| Mn(1) | 4.395 | 4.053 | <u>3.977</u> | 4.519 | 4.167 | <u>4.089</u> |
| Mn(2) | 4.393 | 4.051 | <u>3.975</u> | 4.573 | 4.217 | <u>4.138</u> |
| Mn(3) | 3.411 | <u>3.146</u> | 3.087 | 4.398 | 4.055 | <u>3.979</u> |
| Mn(4) | 3.126 | <u>2.882</u> | 2.828 | 4.458 | 4.111 | <u>4.034</u> |
| Mn(5) | 3.402 | <u>3.137</u> | 3.078 | 3.435 | <u>3.168</u> | 3.108 |
| Mn(6) | 2.979 | <u>2.747</u> | 2.696 | 3.455 | <u>3.185</u> | 3.126 |
| Mn(7) | | | | 3.549 | <u>3.273</u> | 3.211 |
| Mn(8) | | | | 3.392 | <u>3.128</u> | 3.069 |
| Mn(9) | | | | 3.506 | <u>3.233</u> | 3.172 |
| Mn(10) | | | | 3.464 | <u>3.194</u> | 3.134 |
| Mn(11) | | | | <u>2.304</u> | 2.125 | 2.085 |
| Mn(12) | | | | 3.711 | <u>3.422</u> | 3.358 |

[a] The underlined value is the one closest to the actual charge for which it was calculated. The oxidation state of a particular atom can be taken as the nearest whole number to the underlined value.

**Table 4.** Magnetization Relaxation Parameters for Several [Mn$_{12}$] SMM's

| Complex | Charge of [Mn$_{12}$] Moiety | Peak temperature (K) in $c''_M$ at 1kHz | $U_{eff}/k_B$ (K) [a] | $t_0$(s) [a] | Notes |
|---|---|---|---|---|---|
| **1** | 0 | 7.3[b] | 61 | 2.1 x 10$^{-7}$ | Ref. 4,7 |
| **2** | 0 | 5.9 | 64 | 3.4 X 10$^{-9}$ | this work |
| **2a** | 1- | 4.8 | 54 | 2.1 x 10$^{-9}$ | this work |
| **2b** | 1- | 4.6 | 50 | 3.0 x 10$^{-9}$ | this work |
| **2c** | 1- | 4.8 | 53 | 2.3 x 10$^{-9}$ | this work |
| **2d** | 1- | 4.8 | 51 | 3.1 x 10$^{-9}$ | this work |
| **2e** | 2- | 2.9 | 28 | 1.6 x 10$^{-8}$ | this work |
| **2f** | 2- | 2.8 | 27 | 9.0 x 10$^{-9}$ | this work |
| Mn$_{12}$-benzoate-*m*-MPYNN[c] | 1- | 3.9[b] | 50 | 3.0 x 10$^{-10}$ | Ref. 10 |

[a]Estimated from Arrhenius plots using eq 2.  [b]Extrapolated to 1 kHz ac frequency using original data.  [c]*m*-MPYNN: *m*-*N*-methylpyridinium nitronylnitroxide.

**Table 5.** Resonance Field Transitions for [Fe(C$_5$Me$_5$)$_2$][Mn$_{12}$O$_{12}$(O$_2$CC$_6$F$_5$)$_{16}$-(H$_2$O)$_4$]•2H$_2$O (**2a**•2H$_2$O)

| transition $M_s \rightarrow M_s + 1$ | resonance field (T) | | |
|---|---|---|---|
| | 217.23 GHz | 326.26 GHz | 434.24 GHz |
| -21/2 → -19/2 | 0.414 | 4.325 | 8.29 |
| -19/2 →-17/2 | 1.107 | 5.059 | 9.049 |
| -17/2 →-15/2 | 1.894 | 5.863 | 9.835 |
| -15/2 →-13/2 | 2.695 | 6.667 | 10.567 |
| -13/2 →-11/2 | 3.446 | 7.437 | 11.465 |
| -11/2 → -9/2 | 4.324 | 8.359 | |
| -9/2 →-7/2 | 5.064 | 9.066 | |
| -7/2 →-5/2 | 5.857 | | |
| -5/2 →-3/2 | 6.663 | | |
| -3/2 →-1/2 | 7.435 | | |
| -1/2 →1/2 | 8.338 | | |
| 1/2 → 3/2 | 9.069 | | |

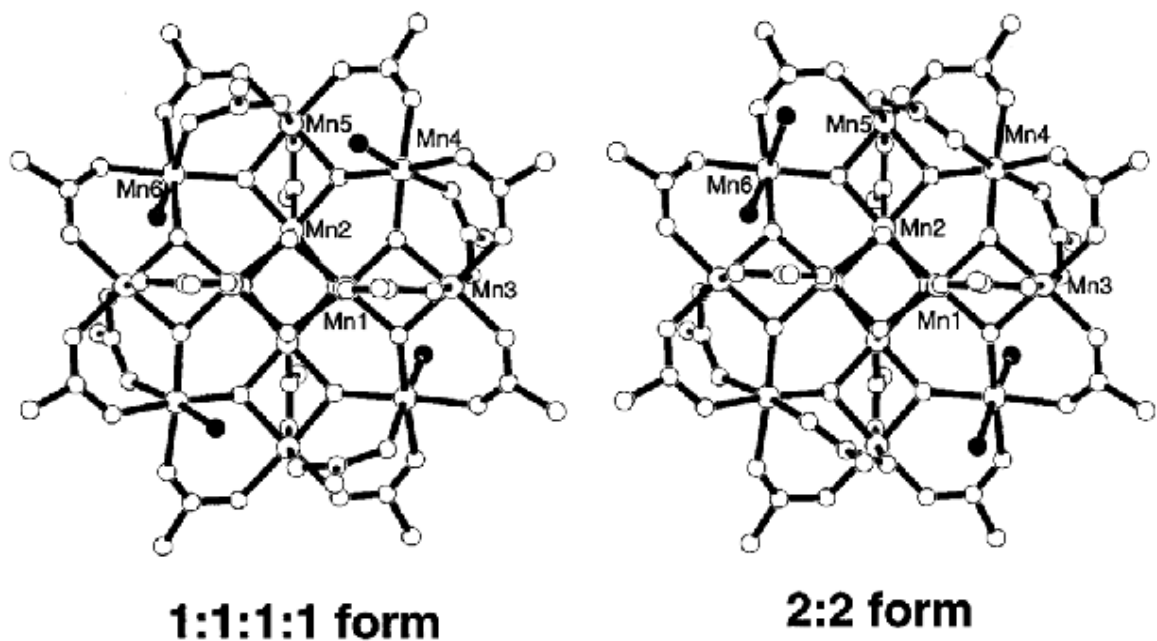

**1:1:1:1 form**   **2:2 form**

**Figure 1.** [Mn$_{12}$O$_{12}$(O$_2$CC)$_{16}$O$_4$] core structures of two water isomers, the 1:1:1:1 form and the 2:2 form, in **2a**•2H$_2$O. The isomers result from the disordering of two carboxylates out of sixteen. Oxygen atoms of water molecules are indicated by filled circles.

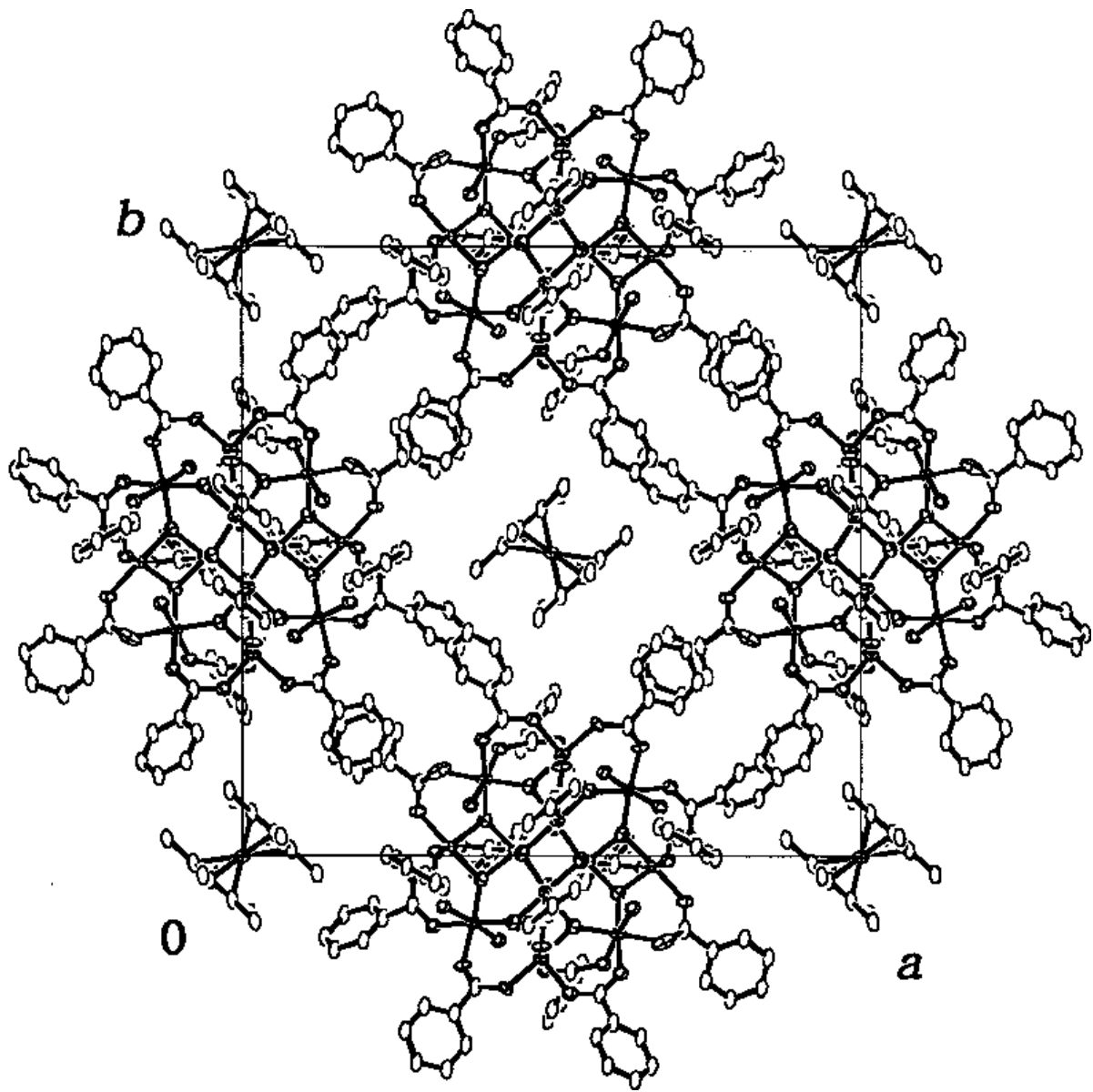

**Figure 2.** Molecular packing of **2a**•2H₂O viewed from the *c*-axis. Hydrogen and fluorine

atoms and water oxygen atoms incorporated into crystal were omitted for clarity.

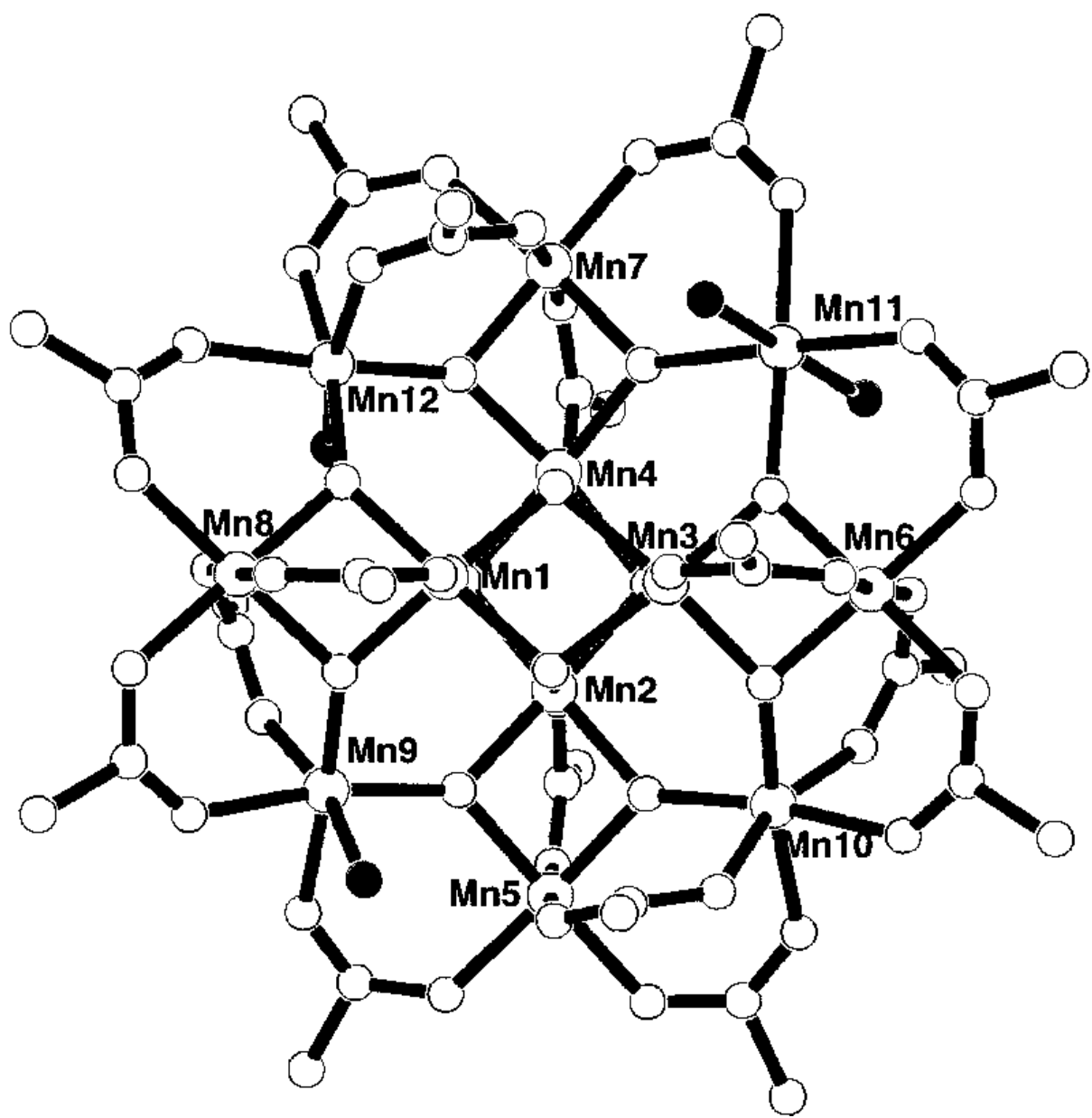

**Figure 3.** The [Mn$_{12}$O$_{12}$(O$_2$CC)$_{16}$O$_4$] core structure in **2c**•2CH$_2$Cl$_2$•C$_6$H$_{14}$ with a 2:1:1 water arrangement. Water oxygen atoms are indicated by filled circles.

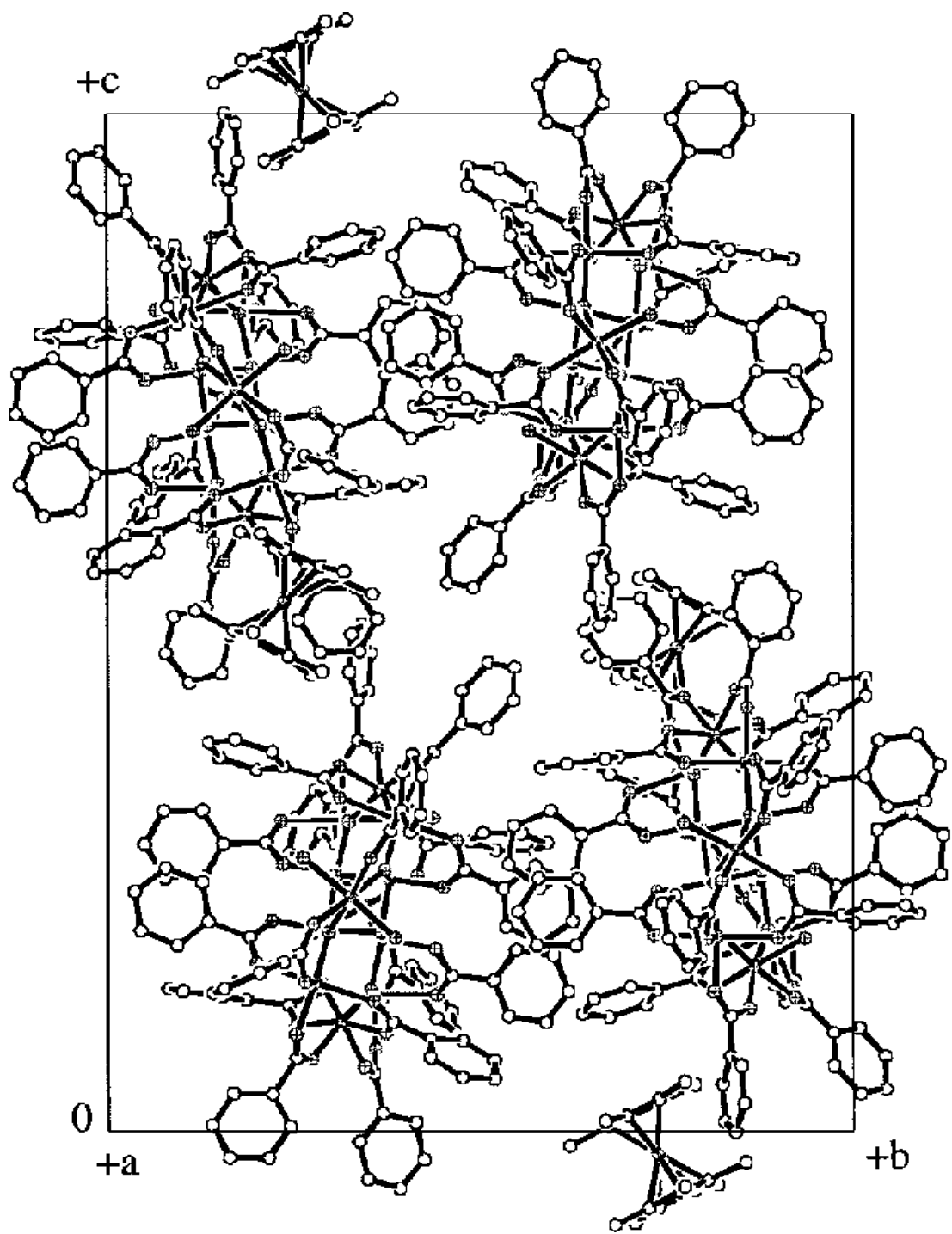

**Figure 4.**

Molecular packing of **2c**•2CH$_2$Cl$_2$•C$_6$H$_{14}$ viewed from the *a*-axis. Hydrogen and fluorine atoms and solvated molecules incorporated into crystal were omitted for clarity. The easy axes of magnetization of each [Mn$_{12}$]$^-$ anion are canted by 6.9° from the *b*-axis.

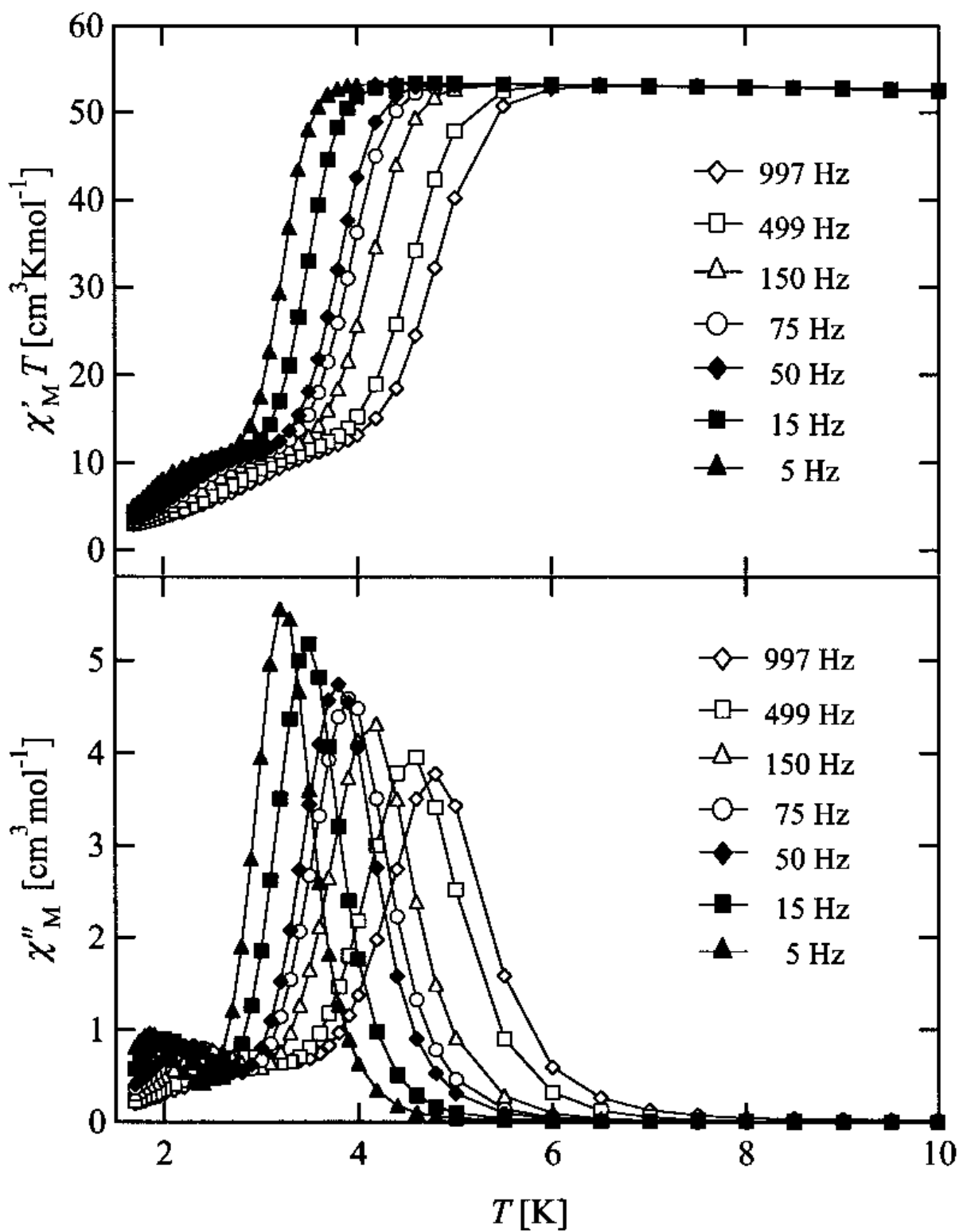

**Figure 5.**     Plots of $\chi'_M T$ vs $T$ (top) and $\chi''_M$ vs $T$ (bottom) for a polycrystalline sample of complex **2a•2H$_2$O** in a 0.1 mT ac field oscillating at the indicated frequencies, where $\chi'_M$ and $\chi''_M$ are the in-phase and the out-of-phase magnetic susceptibility, respectively.

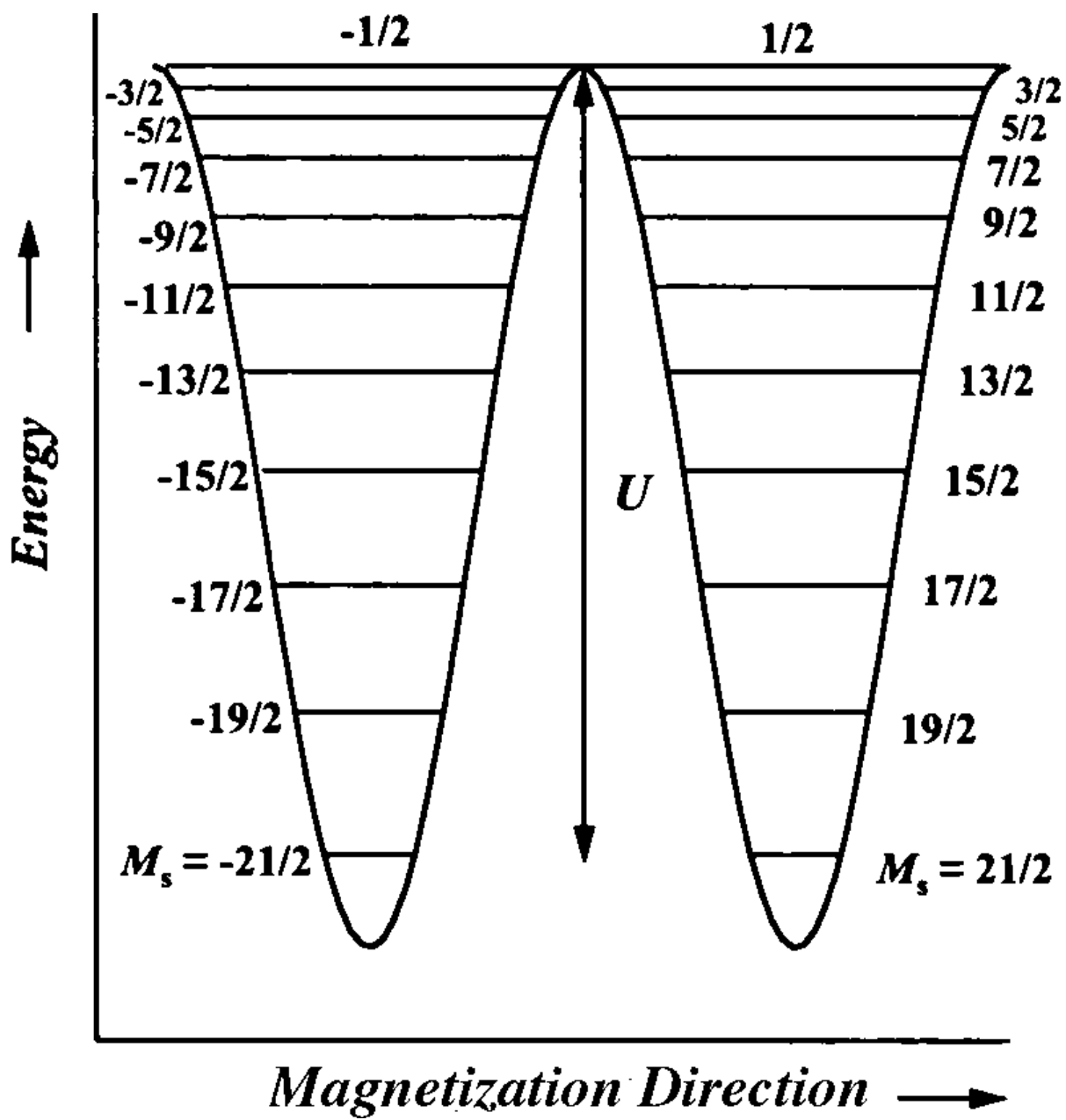

**Figure 6.**     Plot of potential energy vs the magnetization direction for a single molecule with a $S = 21/2$ ground state split by axial zero-field splitting. The potential energy barrier $U$ is given by $110|D|$.

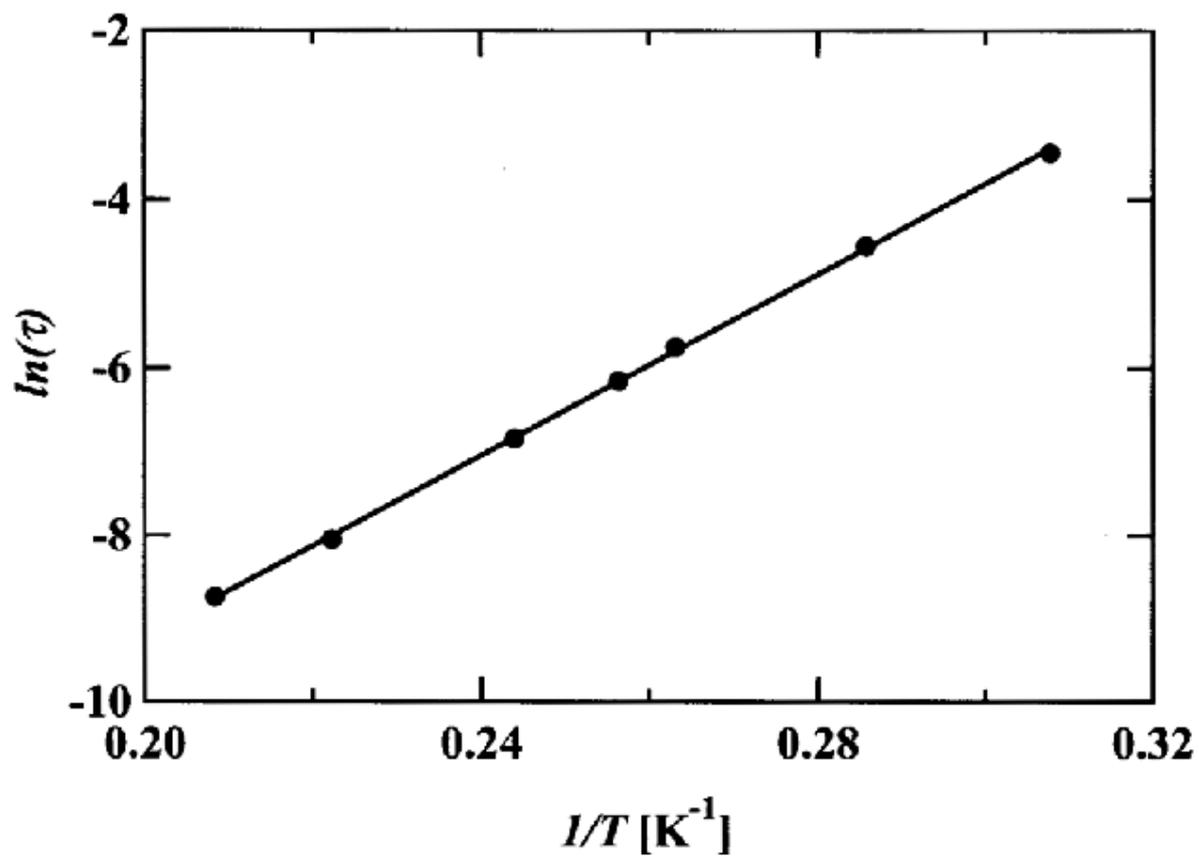

**Figure 7.** Plot of the natural logarithm of the magnetization relaxation rate [ln($\tau$)] vs the inverse of the absolute temperature for complex **2a**•2H$_2$O. The solid line represents a least-squares fit of the data to the Arrhenius equation (see text).

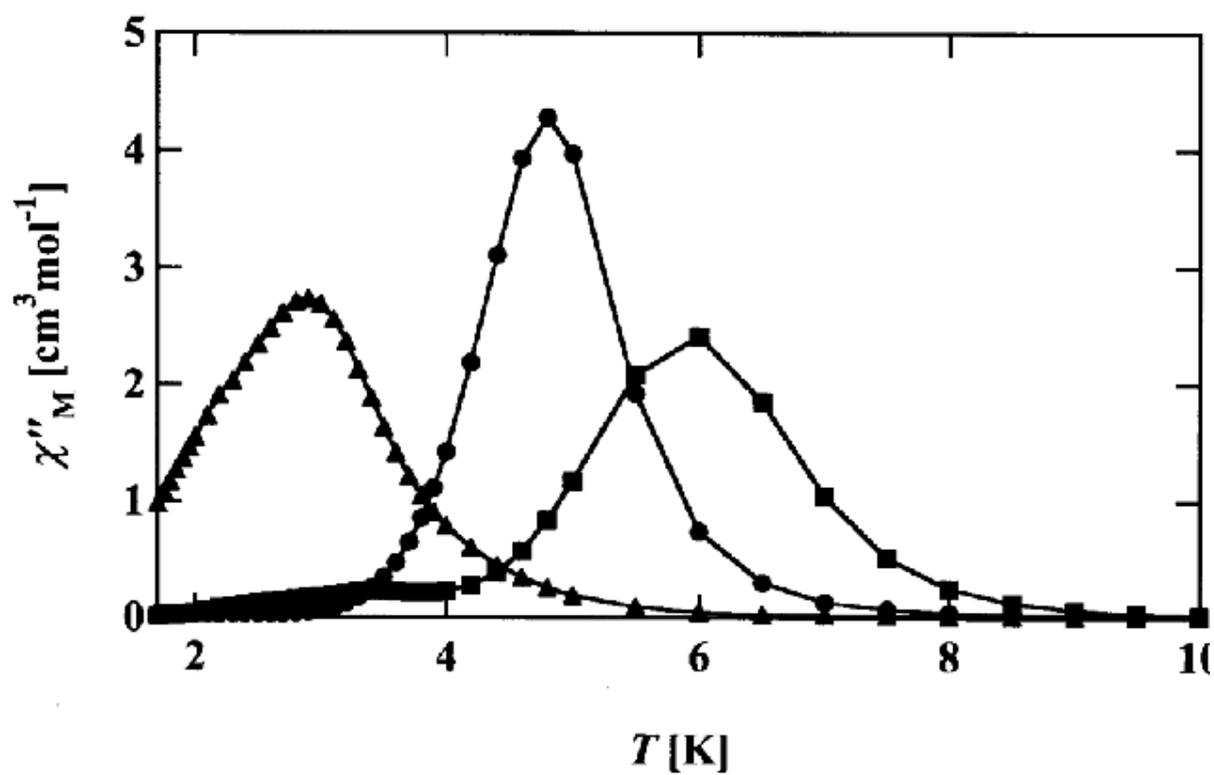

**Figure 8.** Ac out-of-phase signals $(\chi''_M)$ of [Fe(C$_5$Me$_5$)$_2$]$_n$[Mn$_{12}$O$_{12}$(O$_2$CC$_6$F$_5$)$_{16}$(H$_2$O)$_4$] ($n$ = 0: **2** (■), $n$ = 1: **2a** (●), $n$ = 2: **2e** (▲)). The lines are visual guides.

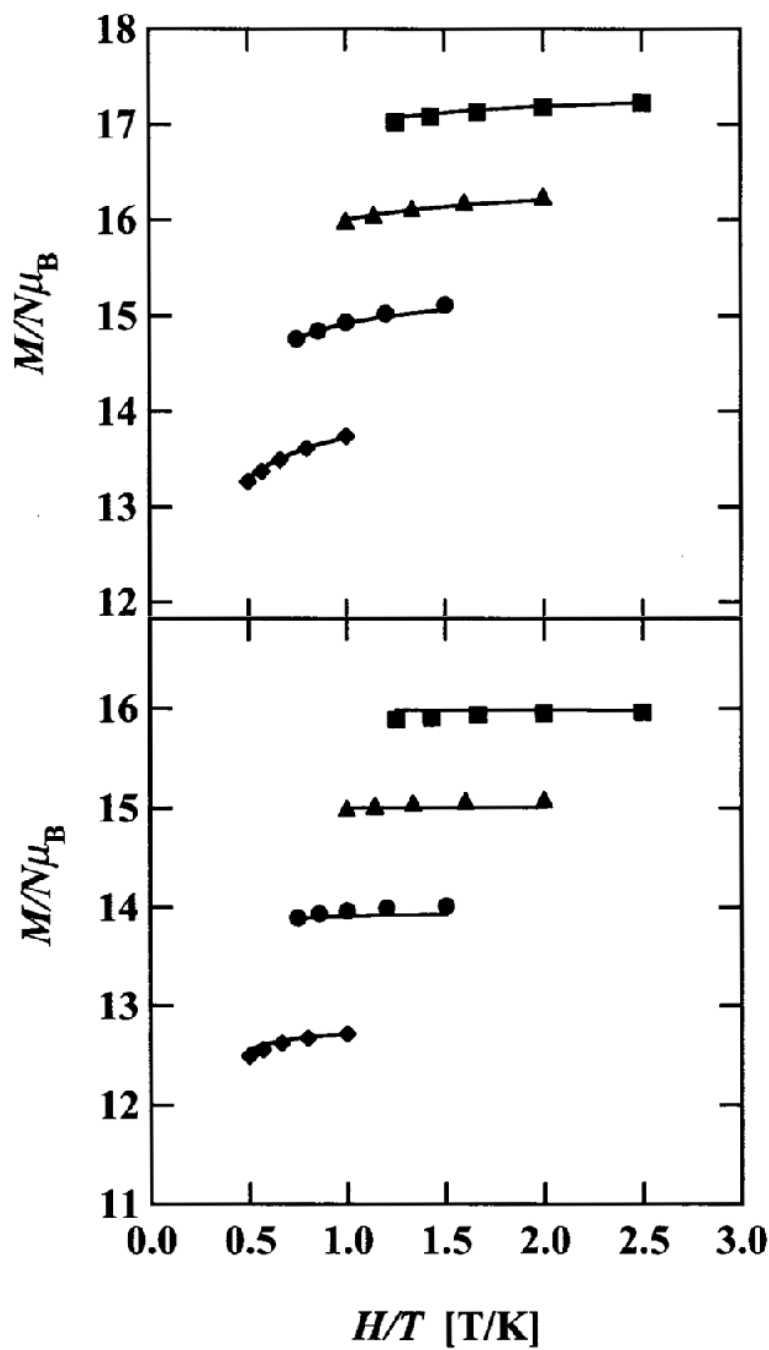

**Figure 9.** Plots of the reduced magnetization $M/(N\mu_B)$ vs the ratio of the external field and the absolute temperature for **2a•2H$_2$O** (top) and **2d** (bottom). The solid lines are fits of the data to an $S = 21/2$ state with $g = 1.92$ and $D = -0.38$ cm$^{-1}$ for **2a•2H$_2$O**, and to an $S = 21/2$ state with $g = 1.89$ and $D = -0.36$ cm$^{-1}$ for **2d**. The contribution of paramagnetic [Fe(C$_5$Me$_5$)$_2$]$^+$ cations ($S = 1/2$, $g_\parallel = 4.43$, $g_\perp = 1.35$) is also included for **2a•2H$_2$O**. Data were measured at 2 (◆), 3 (●), 4 (▲), and 5 T(■).

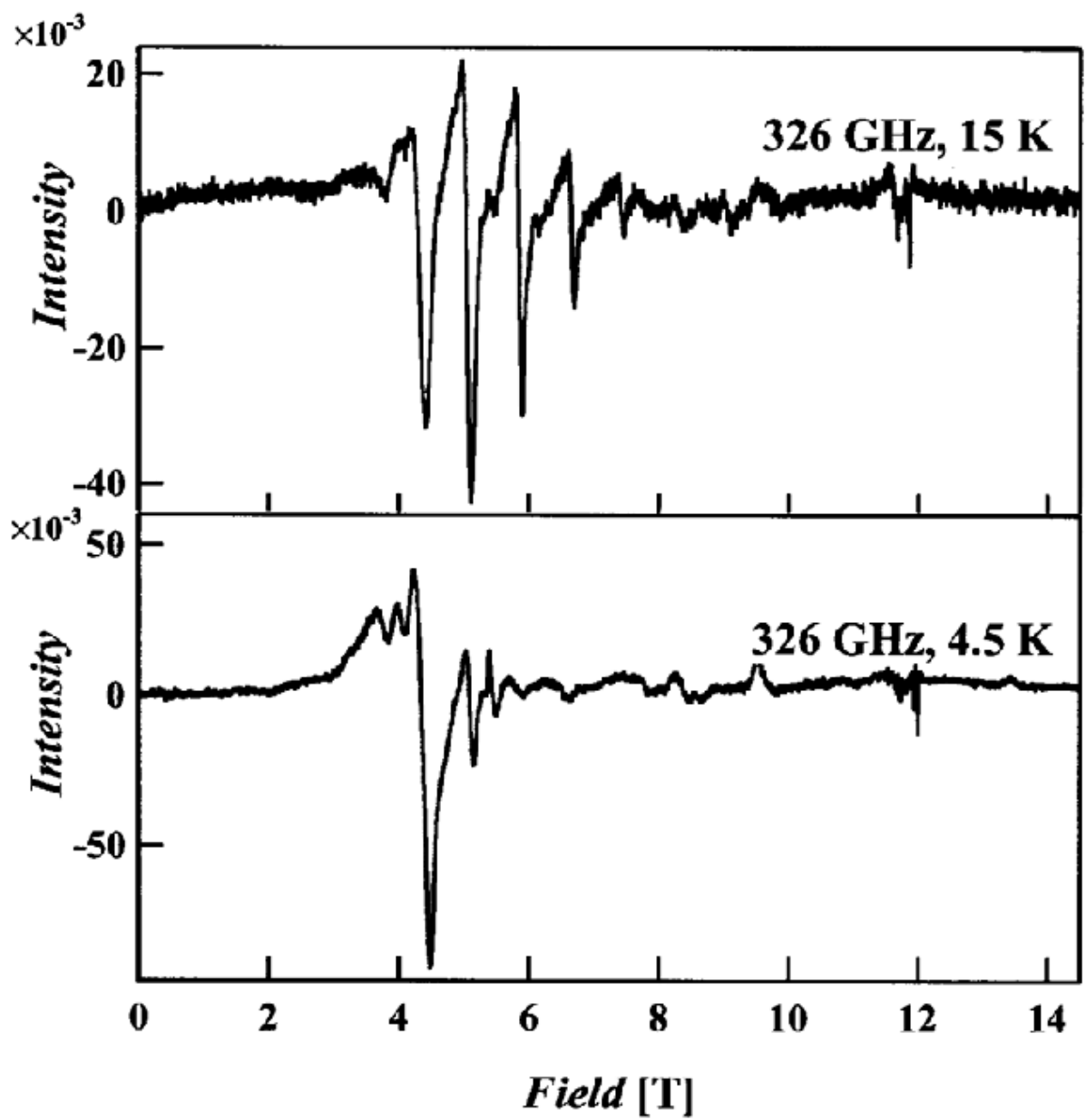

**Figure 10.** HFEPR spectra recorded at 326.26 GHz for a microcrystalline sample of complex **2a**•2H$_2$O that had been oriented in a strong field. The top plot is of the 15 K spectrum, whereas the 4.5 K spectrum is shown in the bottom plot.

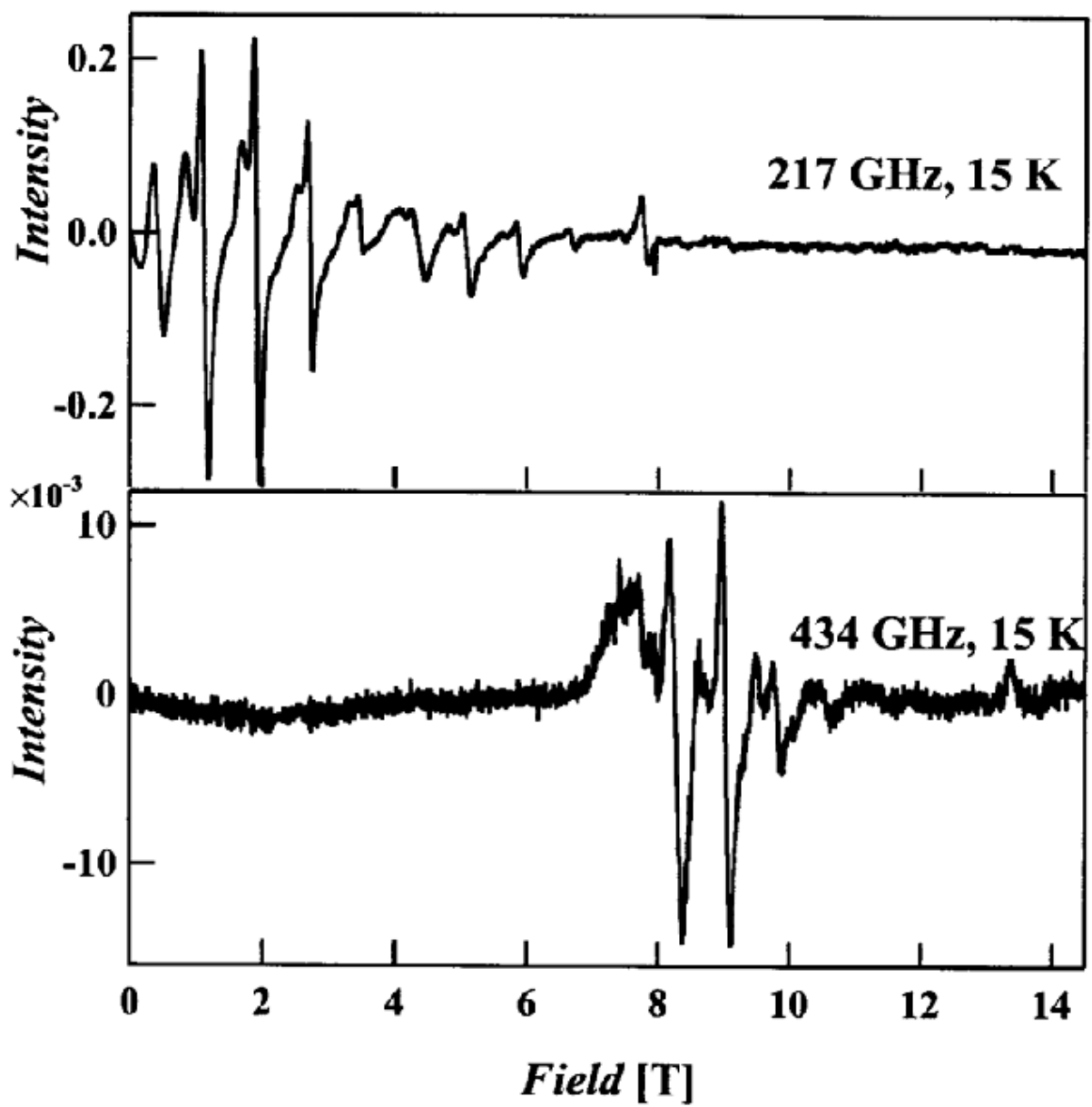

**Figure 11.** HFEPR spectra for a microcrystalline sample of complex **2a**•2H$_2$O recorded at 217.23 GHz (top) and 434.24 GHz (bottom) for temperature 15 K.

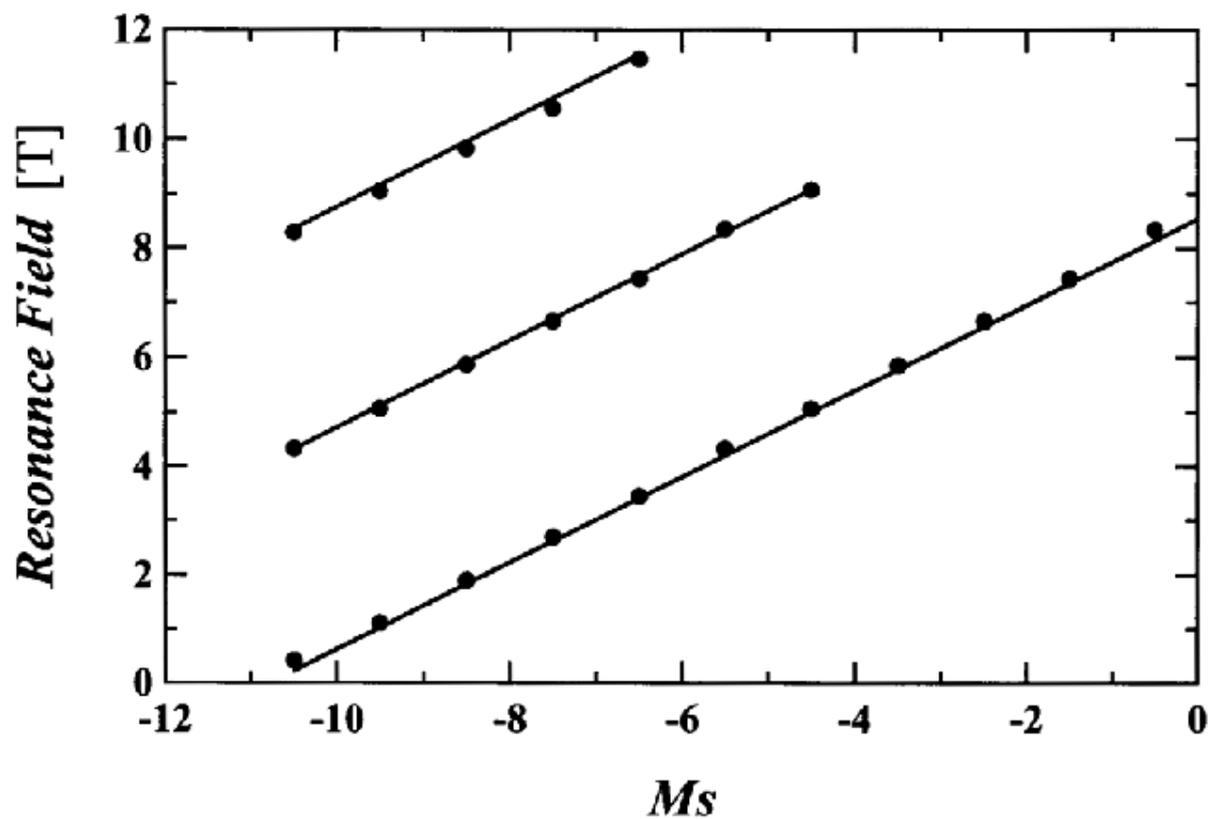

**Figure 12.** Plot of the resonance field vs the $M_s$ number for the HFEPR transitions between the $M_s$ and ($M_s + 1$) zero-field components of the ground state of complex **2a**•2H$_2$O. HFEPR data were measures at 217.23, 326.26, and 434.24 GHz. The solid lines represent a fit of the data to the spin Hamiltonian given in eq 6.

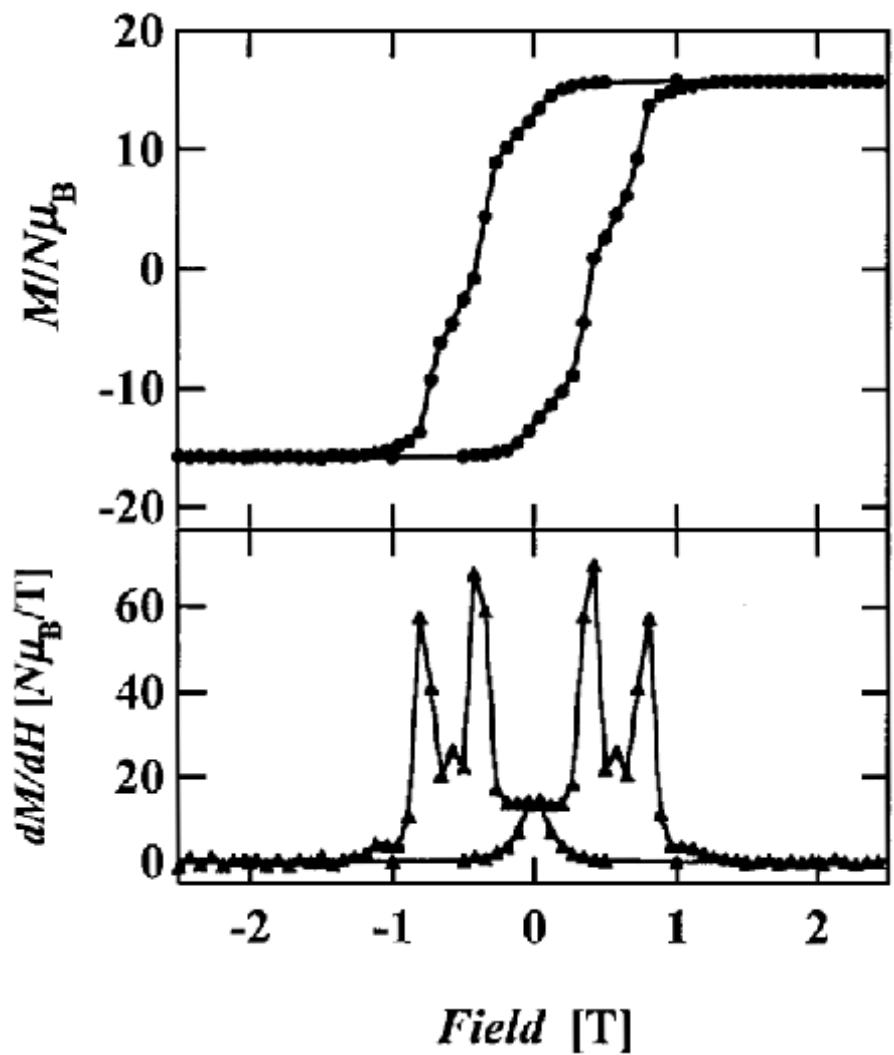

**Figure 13.** Top: Magnetization hysteresis loop measured at 1.85 K for micro crystals of **2c**•2CH$_2$Cl$_2$•C$_6$H$_{14}$. Samples were oriented in eicosane wax matrixes. Bottom: The first derivative of the magnetization hysteresis loop.

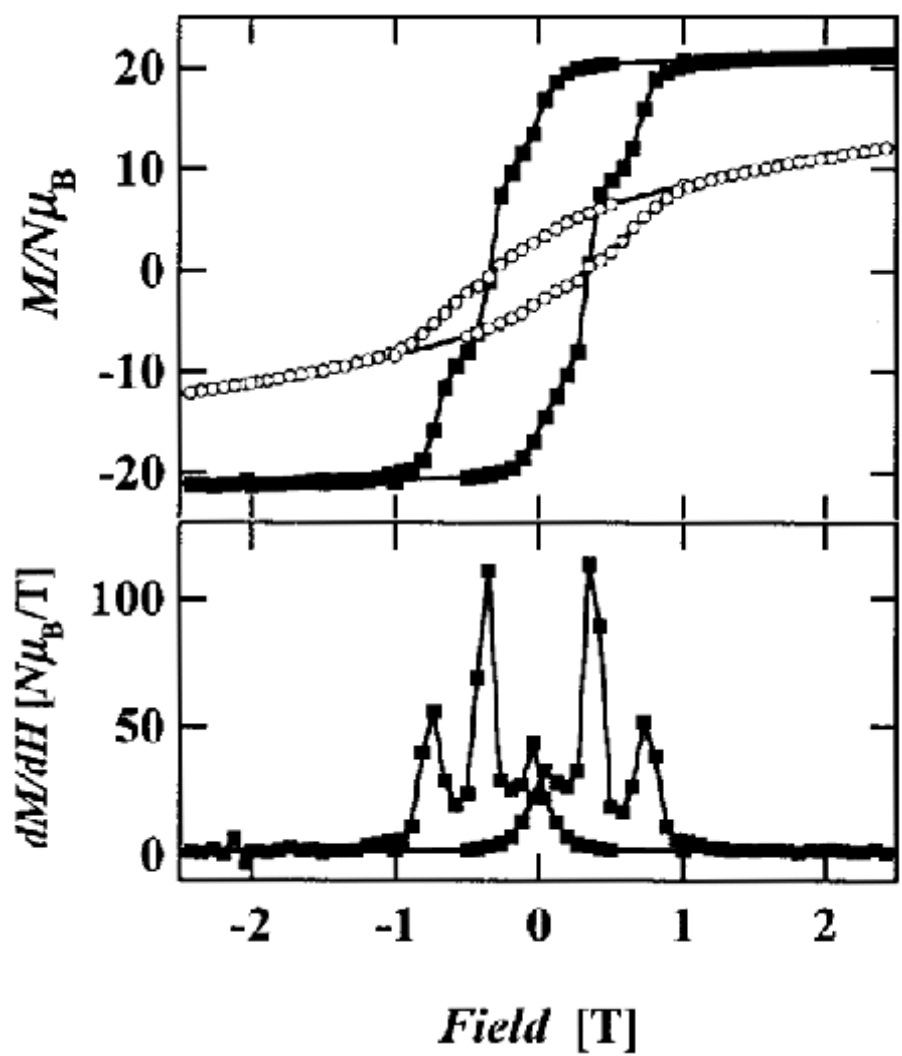

**Figure 14.** Top: Magnetization hysteresis loops measured at 1.85 K for a single crystal of **2a**•2H$_2$O. The crystal was aligned by external magnetic field (5 T) at 318 K, then fixed in eicosane wax and measured (open circles). The sample was then rotated 90° so that the major plane of the crystal could be perpendicular to the applied field, then fixed and measured (filled squares). Bottom: The first derivative of the magnetization hysteresis loop obtained after sample rotation.

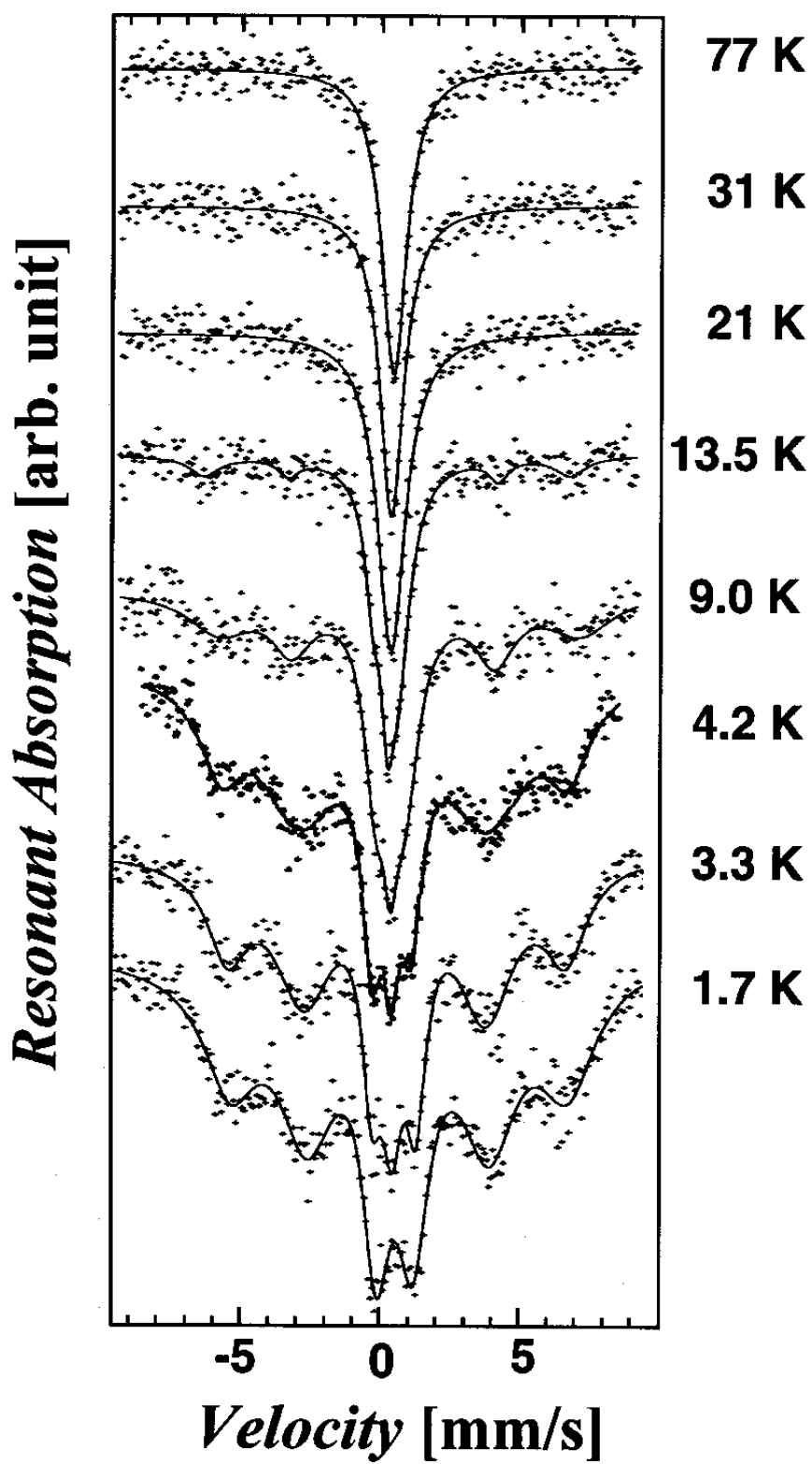

**Figure 15.** Temperature dependence of $^{57}$Fe Mössbauer spectra of 11 % $^{57}$Fe-enriched sample of **2b**.